%% file: ex_article.tex
\documentclass[onefignum,onetabnum]{siamonline171218}

\input{ex_shared}

\ifpdf
\hypersetup{
  pdftitle={A Data-Integrated Framework for Learning Fractional-Order Nonlinear Dynamical Systems},
}
\fi

\myexternaldocument{ex_supplement}

\begin{document}

\maketitle

\begin{abstract}
This paper presents a data-integrated framework for learning the dynamics of fractional-order nonlinear systems in both discrete-time and continuous-time settings. The proposed framework consists of two main steps. In the first step, input-output experiments are designed to generate the necessary datasets for learning the system dynamics, including the fractional order, the drift vector field, and the control vector field. In the second step, these datasets, along with the memory-dependent property of fractional-order systems, are used to estimate the system's fractional order. The drift and control vector fields are then reconstructed using orthonormal basis functions. To validate the proposed approach, the algorithm is applied to four benchmark fractional-order systems. The results confirm the effectiveness of the proposed framework in learning the system dynamics accurately. Finally, the same datasets are used to learn equivalent integer-order models. The numerical comparisons demonstrate that fractional-order models better capture long-range dependencies, highlighting the limitations of integer-order representations.
\end{abstract}



\section{Introduction}\label{section:introduction}
Fractional calculus, which generalizes the concepts of integration and differentiation to non-integer orders~\cite{podlubny1998fractional}, has been extensively used in control theory and dynamical systems due to its accuracy in representing system dynamics, its ability to satisfy additional control requirements, and its capacity to model non-Markovian behavior and long-range dependence~\cite{azar2017fractional}. In recent years, fractional-order systems have been employed to model a wide range of physical and engineering phenomena, including viscoelastic materials~\cite{meral2010fractional}, diffusion waves~\cite{bhrawy2015spectral}, drug administration~\cite{yaghooti2020constrained}, lithium-ion batteries~\cite{wang2020experimental}, power networks~\cite{shalalfeh2016evidence}, and signal processing~\cite{sheng2011fractional}. Correspondingly, various control strategies for fractional-order systems have been proposed, such as fractional PID controllers~\cite{yaghooti2018robust}, fractional CRONE controllers~\cite{wardi2022fractional}, and fractional lead-lag controllers~\cite{abdulkhader2018fractional}. To further enhance control performance in nonlinear systems, several fractional-order extensions of classical methods have been introduced, including sliding mode control~\cite{yaghooti2020adaptive}, model reference adaptive control~\cite{balaska2021direct}, backstepping control~\cite{doostdar2022adrc}, and fuzzy control~\cite{mohammadzadeh2020novel}.

The application of fractional calculus in dynamical systems theory and signal processing has demonstrated improved accuracy in modeling non-Markovian behavior and long-range dependence compared to integer-order models~\cite{shalalfeh2016evidence,ghorbani2018gene}. Despite these advantages, learning the dynamics of fractional-order systems remains a challenging task. Over the years, several learning and system identification methods for such systems have been proposed. For instance, the work in~\cite{idiou2014linear} introduces a learning-based approach for fractional-order linear systems using time-domain data. This method applies a recursive least squares algorithm to an ARX structure, which is derived from the fractional-order differential equations using adjustable fractional-order differentiators. The approach in~\cite{hartley2003fractional} employs continuous-order distributions to characterize and identify fractional-order systems. The method proposed in~\cite{shah2019closed} estimates model parameters by minimizing the sum of squared errors using a genetic algorithm. Similarly, \cite{chaudhary2018novel} presents a generalization of the Volterra LMS algorithm adapted to fractional-order dynamics. In a more recent contribution, \cite{gupta2019learning} proposes an iterative framework for learning network parameters and predicting the states of discrete-time fractional-order systems. Finally,\cite{yaghooti2023inferring} introduces a dynamic inference algorithm specifically designed for discrete-time settings, and\cite{zhang2025sampling} provides a sample complexity analysis of the method proposed in~\cite{yaghooti2023inferring}.

In this paper, we present a data-integrated framework for learning the dynamics of control-affine fractional-order systems in both discrete-time and continuous-time settings. The proposed algorithms assume that the systems are control-affine and that the number of state variables is known. However, the drift vector field, the control vector field, and the system's fractional order are assumed to be unknown.
The framework consists of two main steps. In the first step, a series of experiments is designed to generate datasets for learning the system dynamics. To construct the first dataset, a fixed initial condition is applied, and a variety of control inputs are introduced while recording the resulting state trajectories over two consecutive time steps. This procedure is repeated for multiple initial conditions to capture a broader range of system behavior. The second and third datasets are generated using the same procedure, with the initial conditions now set to the system responses resulting from the initial conditions used in the first dataset. This approach captures the memory-dependent nature of fractional-order systems across sequential experiments.
The second step of the algorithm involves estimating the fractional order of the system and reconstructing the drift and control vector fields. The fractional order is estimated by leveraging the memory-dependent structure inherent to fractional-order dynamics. The drift and control vector fields are then approximated using orthonormal basis functions, and the corresponding coefficients are identified by solving an optimization problem that minimizes the mean squared error between the experimentally observed values and the values predicted by the learned model.
Finally, to demonstrate the advantages of fractional-order modeling, a standard system identification method is applied to the same datasets to obtain an integer-order model. The time responses of the fractional-order and integer-order systems are then compared. The simulation results indicate that integer-order models fail to capture the memory-dependent behavior present in the system, whereas fractional-order models are able to accurately learn and represent this property.

The remainder of the paper is organized as follows. Section~\ref{section:preliminary_concepts} introduces the definitions of the fractional-order derivative, the fractional-order difference operator, and both continuous-time and discrete-time fractional-order systems. Section~\ref{section:problem_statement} presents the problem formulation. Section~\ref{section:proposed_algorithm} describes the proposed framework for learning the dynamics of fractional-order nonlinear systems. Section~\ref{section:simulation} reports the simulation results that validate the proposed methodology. Finally, Section~\ref{section:conclusion} concludes the paper and outlines directions for future research.

\section{Preliminary Concepts}\label{section:preliminary_concepts}
Since Leibniz first introduced the concept of fractional-order derivatives and integrals, numerous definitions for fractional-order integrals, derivatives, and difference operators have been developed. In this section, we present the Caputo fractional derivative, the Gr\"unwald-Letnikov operator, and the formulations of fractional-order continuous-time and discrete-time systems.
\begin{definition}[Caputo Fractional Derivative~\cite{podlubny1998fractional}]
The Caputo fractional derivative of order $\alpha \in \bR$ for a function $f: \bR \rightarrow \bR$ is defined as
\begin{align}\label{eq:caputo_definition}
    {}^{C}_{0}D^{\alpha}_{t}f(t)= 
    \begin{cases}
        \displaystyle \frac{1}{\Gamma(k-\alpha)} \int_{0}^{t} \frac{f^{(k)}(\tau)}{(t-\tau)^{\alpha+1-k}}\, d\tau, & \text{if } k-1<\alpha<k, \\
        \displaystyle \frac{d^{k}}{dt^{k}}f(t), & \text{if } \alpha=k,
    \end{cases}
\end{align}
where $k$ is the smallest integer such that $k - 1 < \alpha \leq k$.
\end{definition}

\begin{notation}
For brevity, the operator ${}^{C}_{0}D^{\alpha}_{t}$ is denoted by $D^{\alpha}_{t}$ throughout the remainder of this paper.
\end{notation}

\begin{definition}[Continuous-time Fractional-Order Systems~\cite{podlubny1998fractional}]
A continuous-time, frac- tional-order, control-affine nonlinear system is described using the Caputo derivative as follows:
\begin{align}\label{eq:general_continuous_nonlinear_system}
    D^\alpha_t x(t) = f(x(t)) + g(x(t)) u(t),
\end{align}
where $x \in \bR^n$ is the state vector, $u(t): \bR \rightarrow \bR^m$ is the system input, $f: \bR^n \rightarrow \bR^n$ and $g: \bR^n \rightarrow \bR^{n \times m}$ are nonlinear functions, and $\alpha = [\alpha_1, \alpha_2, \dots, \alpha_n]^\intercal \in \bR^n$ denotes the vector of fractional orders. The operator $D^\alpha_t$ is defined as
\begin{align*}
    D^\alpha_t = \operatorname{diag} \left( D_t^{\alpha_1}, D_t^{\alpha_2}, \dots, D_t^{\alpha_n} \right),
\end{align*}
where $\operatorname{diag}(\cdot)$ denotes a diagonal matrix with the specified elements on its main diagonal.
\end{definition}

The algorithms proposed in this paper are data-driven, and discretization of the system is essential for data collection and dynamic reconstruction. Proposition~\ref{proposition:discretization} presents the discretization procedure for continuous-time fractional-order systems.
\begin{proposition}[Discretization of a Fractional-Order System~\cite{mendes2019numerical}]\label{proposition:discretization}
For the fractional-order system described in~\eqref{eq:general_continuous_nonlinear_system}, a discretization method for obtaining numerical solutions with step size~$h$ and infinite memory initialized at $x(0)$ is given by
\begin{align}\label{eq:numerical_solution_FIVP}
    x(k+1) = h^{\alpha} \left(f(x(k))+ g(x(k)) u(k)\right)-\sum_{j=1}^{k+1} \psi(\alpha,j) x(k-j+1) +\bigg(I_{n \times n}+\sum_{j=1}^{k+1} \psi(\alpha,j)\bigg)x(0),
\end{align}
where $I_{n\times n}$ is the identity matrix, and $h^{\alpha}$ and $\psi(\alpha,j)$ are diagonal matrices defined as
\begin{align}
    h^{\alpha} &= \operatorname{diag}\left(h^{\alpha_1}, h^{\alpha_2}, \ldots, h^{\alpha_n} \right), \\
    \psi(\alpha,j) &= \operatorname{diag}\left( \psi(\alpha_1,j), \psi(\alpha_2,j), \ldots, \psi(\alpha_n,j) \right), \label{eq:psi_definition}
\end{align}
with
\begin{align*}
    \psi(\alpha_i,j) = \frac{\Gamma(j - \alpha_i)}{\Gamma(-\alpha_i)\Gamma(j + 1)}, \quad i = 1, 2, \ldots, n,
\end{align*}
and $\Gamma(\cdot)$ denotes the Gamma function, defined as $\Gamma(z) = \int_{0}^{\infty}t^{z-1}e^{-t} dt$.
\end{proposition}

\begin{definition}[Gr\"unwald-Letnikov Difference Operator~\cite{podlubny1998fractional}]
The fractional-order Gr\"unw- ald-Letnikov difference operator is defined as
\begin{align}\label{eq:grunwald_letnikov_def}
    \Delta^{\alpha} x(k) = \sum_{j=0}^{k} \psi(\alpha,j) x(k-j),
\end{align}
where $x(k) \in \bR^n$, and $\alpha = [\alpha_1, \alpha_2, \dots, \alpha_n]^\intercal \in \bR^n$ denotes the vector of fractional orders. The operator $\Delta^{\alpha}$ is defined componentwise as
\begin{align}
    \Delta^{\alpha}=\operatorname{diag}\left( \Delta^{\alpha_1}, \Delta^{\alpha_2}, \dots, \Delta^{\alpha_n} \right),
\end{align}
and $\psi(\alpha,j)\in\bR^{n\times n}$ is a diagonal matrix as defined in~\eqref{eq:psi_definition}.
\end{definition}

\begin{definition}[Discrete-time Fractional-Order Systems~\cite{sadeghian2013general}]
A discrete-time, fractional-order, control-affine nonlinear system is defined as
\begin{align}\label{eq:general_discrete_nonlinear_system_1}
    \Delta^{\alpha} x(k+1) = f(x(k)) + g(x(k)) u(k),
\end{align}
where $x(k) \in \bR^n$ is the system state vector, $u(k) \in \bR^m$ is the control input, and $f:\bR^n \rightarrow \bR^n$ and $g: \bR^n \rightarrow \bR^{n \times m}$ are nonlinear functions.
\end{definition}

By applying the definition of the Gr\"unwald-Letnikov difference operator, the system in~\eqref{eq:general_discrete_nonlinear_system_1} can be rewritten as
\begin{align}\label{eq:general_discrete_nonlinear_system_2}
    x(k+1) & = \Delta^{\alpha} x(k+1) - \sum_{j=1}^{k+1} \psi(\alpha,j) x(k+1-j) \nonumber \\
    & = f(x(k)) + g(x(k)) u(k) - \sum_{j=1}^{k+1} \psi(\alpha,j) x(k+1-j).
\end{align}
The summation term on the right-hand side of~\eqref{eq:general_discrete_nonlinear_system_2} reflects the memory-dependent nature of fractional-order dynamical systems. This property enables such systems to effectively model non-Markovian behavior and long-range dependencies. In contrast, for integer-order systems, this term vanishes, and the next state depends only on the current state.

\section{Problem Statement}\label{section:problem_statement}
In this section, we formulate two problems related to learning the dynamics of continuous-time and discrete-time systems. These two cases are treated separately due to the distinct mathematical challenges associated with each setting. In particular, the discretization of fractional-order continuous-time systems is considerably more complex than that of integer-order systems. This complexity arises from the nonlocal nature of fractional derivatives, which introduce memory effects that must be preserved throughout the discretization process. As a result, specialized numerical techniques are required to obtain an accurate discrete-time representation.

\subsection{Continuous-time Systems}\label{section:problem_statement_continuous}
Consider the following control-affine, continuous-time fractional-order nonlinear system:
\begin{align}\label{eq:continuous_nonlinear_system_1}
    D_t^{\alpha} x(t) = f(x(t)) + g(x(t)) u(t),
\end{align}
where $x(t) \in \cX \subseteq \bR^n$ is the system state vector, $\alpha=\alpha_1,\alpha_2,\dots,\alpha_n]^\intercal\in\bR^n$ denotes the vector of fractional orders, and $u(t) \in \bR^m$ is the system input. The functions $f:\cX\rightarrow\bR^n$ and $g:\cX\rightarrow\bR^{n\times m}$ are unknown nonlinear mappings, defined component-wise as follows:
\begin{align}
    f(x) & = \begin{bmatrix} f_1(x) & f_2(x) & \cdots & f_n(x) \end{bmatrix}^\intercal, \\
    g(x) & = \begin{bmatrix} g_{1,1}(x) & g_{1,2}(x) & \cdots & g_{1,m}(x) \\
    \vdots & \vdots & \ddots & \vdots \\
    g_{n,1}(x) & g_{n,2}(x) & \cdots & g_{n,m}(x)
    \end{bmatrix},
\end{align}
where $f_i(\cdot):\cX\rightarrow\bR$ and $g_{i,j}:\cX\rightarrow\bR$ for $i=1,2,\dots,n$ and $j=1,2,\dots,m$.

The objective of this paper is to develop a data-integrated framework that generates data and uses it to learn the dynamics of the system in~\eqref{eq:continuous_nonlinear_system_1}. To this end, we derive an equivalent discrete-time representation of~\eqref{eq:continuous_nonlinear_system_1}. Using the numerical approximation provided in~\eqref{eq:numerical_solution_FIVP}, we obtain the sampled-data system under Zero-Order Hold (ZOH) control as
\begin{align}\label{eq:continuous_nonlinear_system_2}
    x(k+1) = h^{\alpha} \left(f(x(k)) + g(x(k)) u(k)\right)
    - \sum_{j=1}^{k+1} \psi(\alpha, j) x(k - j + 1)
    + \bigg(I_{n \times n} + \sum_{j=1}^{k+1} \psi(\alpha, j)\bigg) x(0).
\end{align}

\begin{problem}\label{Problem_continuous}
Consider the system given in~\eqref{eq:continuous_nonlinear_system_1}, where the state variables are measurable. Assume that the fractional order~$\alpha$, the drift vector field, and the control vector field are all unknown. The objective is to develop an algorithm that estimates the fractional order~$\alpha$ as well as the functions~$f_i(\cdot)$ and~$g_{i,j}(\cdot)$ for all $i=1,2,\dots,n$ and $j=1,2,\dots,m$.
\end{problem}

\subsection{Discrete-time Systems}\label{section:problem_statement_discrete}
Consider the following control-affine, discrete-time fracti-onal-order nonlinear system:
\begin{align}\label{eq:discrete_nonlinear_system_1}
\Delta^{\alpha}x(k+1) = f(x(k)) + g(x(k))u(k),
\end{align}
where $\alpha=[\alpha_1,\alpha_2,\dots,\alpha_n]^\intercal\in\bR^n$ denotes the vector of fractional orders, and $x(k)\in\cX\subseteq\bR^n$ is the system state vector. The input $u(k) \in \bR^m$ is the control input, and the functions $f:\cX\rightarrow\bR^n$ and $g:\cX\rightarrow\bR^{n\times m}$ represent the drift-vector and control-vector fields, defined component-wise as follows:
\begin{align}
    f(x(k)) & = \begin{bmatrix} f_1(x(k)) & f_2(x(k)) & \cdots & f_n(x(k)) \end{bmatrix}^\intercal, \\
    g(x(k)) & = \begin{bmatrix} g_{1,1}(x(k)) & g_{1,2}(x(k)) & \cdots & g_{1,m}(x(k)) \\
    \vdots & \vdots & \ddots & \vdots \\
    g_{n,1}(x(k)) & g_{n,2}(x(k)) & \cdots & g_{n,m}(x(k))
    \end{bmatrix},
\end{align}
where $f_i:\cX \rightarrow \bR$ and $g_{i,j}:\cX\rightarrow\bR$ for $i=1,2,\dots,n$ and $j=1,2,\dots,m$ are unknown nonlinear functions. 

\noindent Following the same procedure used to derive~\eqref{eq:general_discrete_nonlinear_system_2}, we can rewrite the system in~\eqref{eq:discrete_nonlinear_system_1} as:
\begin{align}\label{eq:discrete_nonlinear_system_2}
x(k+1) = f(x(k)) + g(x(k)) u(k) - \sum_{j=1}^{k+1} \psi(\alpha, j) x(k+1-j).
\end{align}
The discrete-time analogue of Problem~\eqref{Problem_continuous} is formulated as follows.
\begin{problem}\label{Problem_discrete}
Consider the system in~\eqref{eq:discrete_nonlinear_system_1}, where the state variables are measurable. Assume that the system's fractional order~$\alpha$, the drift vector field, and the control vector field are all unknown. The objective is to develop an algorithm that estimates the fractional order~$\alpha$ as well as the functions~$f_i(\cdot)$ and~$g_{i,j}(\cdot)$ for all $i = 1, 2, \dots, n$ and $j = 1, 2, \dots, m$.
\end{problem}

\section{Proposed Learning Algorithms}\label{section:proposed_algorithm}
In this section, we propose two data-integrated fra- meworks for learning the dynamics of fractional-order nonlinear systems, as formulated in Problems~\eqref{Problem_continuous} and~\eqref{Problem_discrete}. The first framework, referred to as LCF (Learning Continuous-time Fractional-order dynamics), is designed for continuous-time systems. The second framework, referred to as LDF (Learning Discrete-time Fractional-order dynamics), addresses discrete-time systems. Each framework consists of two main steps. In the first step, a set of input-output experiments is designed to generate the datasets required to capture the system's input-response behavior. In the second step, the collected data are used to formulate the problem of learning the system dynamics, including the fractional order, the drift vector field, and the control vector field, as a linear regression task.

\subsection{LCF: Learning Continuous-time Fractional-order dynamics}
In this subsection, we present the Learning Continuous-time Fractional-order dynamics (LCF) method, as detailed in Algorithm~\ref{algorithm:continuous-systems}. This method is developed to infer the dynamics of nonlinear systems governed by fractional-order continuous-time equations. It assumes that the system is control-affine, with a known number of state variables, while the drift vector field, control vector field, and the vector of fractional orders are all unknown. The LCF framework consists of two main stages. In the first stage, a set of input-output experiments is designed to generate datasets that reflect the system's memory-dependent behavior. In the second stage, the collected data are used to estimate the fractional order and reconstruct the drift and control vector fields by solving regression problems over a selected orthonormal basis. The LCF method is specifically designed to address the challenges associated with learning from continuous-time fractional-order systems, particularly those arising from their nonlocal and long-range dependent dynamics.

\begin{notation}
$x(0)^{(i)}$ denotes the $i$-th initial condition used in the experiments. 
$u(0)^{(i,j)}$ and $u(1)^{(i,j)}$ represent the $j$-th control inputs corresponding to the $i$-th initial condition at time steps $k=0$ and $k=1$, respectively. 
$x(1)^{(i,j)}$ and $x(2)^{(i,j)}$ are the resulting system states after applying $u(0)^{(i,j)}$ and $u(1)^{(i,j)}$, starting from the initial condition $x(0)^{(i)}$. 
${\Tilde{x}(2)}^{(i,j)}$ denotes the system state after applying $u(1)^{(i,j)}$ with $x(1)^{(i,j)}$ as the initial condition. 
In the case of discrete-time systems, ${\Tilde{x}(3)}^{(i,j)}$ represents the system state after applying the control input $u(2)^{(i,j)}$ with $x(2)^{(i,j)}$ as the initial condition.
\end{notation}

\begin{algorithm}
\caption{Learning Continuous-Time Fractional-Order Dynamics (LCF)}
\label{algorithm:continuous-systems}
\begin{algorithmic}[1]

\STATE {\textbf{Input:}
\begin{itemize}[leftmargin=*]
  \item $\{x(0)^{(i)}\}_{i=1}^M$: A set of $M$ initial conditions uniformly sampled from $\cX \subseteq \mathbb{R}^n$.
  \item $\{U(0)^{(i,j)}\}_{i=1,j=1}^{M,N}$ and $\{U(1)^{(i,j)}\}_{i=1,j=1}^{M,N}$: Two sets of randomly chosen system inputs, used to excite the system over multiple trials.
  \item $\phi_i^{(1)}, \ldots, \phi_i^{(L)}$, $i = 1, \ldots, n$: A collection of orthonormal basis functions with $L$ terms in the truncated series to approximate the unknown functions.
\end{itemize}
}

\STATE \textbf{Output:} Functions $f(\cdot)$, $g(\cdot)$, and $\alpha$.

\FOR{$\ell = 1, \ldots, m$}
    \STATE Set inputs for the $\ell$-th element only, by assigning $\{u_{\ell}(0)^{(i,j)}\} = \{U_{\ell}(0)^{(i,j)}\}$ and $\{u_{\ell}(1)^{(i,j)}\} = \{U_{\ell}(1)^{(i,j)}\}$. For all other elements $q \neq \ell$, set $\{u_{q}(0)^{(i,j)}\} = \{u_{q}(1)^{(i,j)}\} = 0$.
    

    \STATE \textbf{Data Generation:} From $t=0$ to $t=1$, run the system with initial conditions $\{x(0)^{(i)}\}$ and inputs $\{u(0)^{(i,j)}\}$; record the resulting states $\{x(1)^{(i,j)}\}$. Then, from $t=1$ to $t=2$, run the system with inputs $\{u(1)^{(i,j)}\}$; record $\{x(2)^{(i,j)}\}.$
    

    \STATE {\textbf{System's Order Determination:} Reinitialize the system at $t=1$ using the states $\{x(1)^{(i,j)}\}$ from the previous step. Then, from $t=1$ to $t=2$, run the system with inputs $\{u(1)^{(i,j)}\}$; record the resulting states as $\{\Tilde{x}(2)^{(i,j)}\}$.}

    
    \IF{$\ell = 1$}
        \STATE Determine the system's order $\alpha$ of the system by solving the following equation:
        \[
        x(2)^{(i,j)} - \Tilde{x}(2)^{(i,j)} = (I_{n \times n} + \psi(\alpha, 1))(x(0)^{(i)} - x(1)^{(i,j)}).
        \]
    \ENDIF

    \STATE \textbf{Reconstruction of the control-vector and drift-vector fields:} 
    \begin{itemize} 
    \item Compute $y^{(i,j)} = h^{-\alpha}(x(1)^{(i,j)} - x(1)^{(i,0)})$ for $i = 1, \ldots, M$, $j = 1, \ldots, N$.
    \item Form the matrix $Y = [\,y^{(1,1)}, \ldots, y^{(M,1)}, \ldots, y^{(1,N)}, \ldots, y^{(M,N)}\,]^\intercal$.
    \item Define $\phi^{(i,j)}$ as:
    \[
    \phi^{(i,j)} =
    \begin{bmatrix}
        \phi_1(x(0)^{(i)}) & \mathbf{0} & \cdots & \mathbf{0} \\
        \mathbf{0} & \phi_2(x(0)^{(i)}) & \cdots & \mathbf{0} \\
        \vdots & \vdots & \ddots & \vdots \\
        \mathbf{0} & \mathbf{0} & \cdots & \phi_n(x(0)^{(i)})
    \end{bmatrix}
    \,\Delta u(0)^{(i,j)},
    \]
    where $\Delta u(0)^{(i,j)} = u(0)^{(i,j)} - u(0)^{(i,0)}$.
    \item Let $\Phi = [\,\phi^{(1,1)}, \ldots, \phi^{(M,1)}, \ldots, \phi^{(1,N)}, \ldots, \phi^{(M,N)}\,]^\intercal$.
    \item Solve for $B = (\Phi^\intercal \Phi)^{-1} \Phi^\intercal Y$.
    \item Define $g_{\ell} = [\,\beta_1^\intercal \phi_1, \ldots, \beta_n^\intercal \phi_n\,]^\intercal$.
    \end{itemize}
\ENDFOR

\STATE Return $g = [\,g_1, \ldots, g_m]$ and $f(x(0)^{(i)}) = -\,g(x(0)^{(i)})\,u(0)^{(i,1)} + h^{-\alpha}\bigl(x(1)^{(i,1)} - x(0)^{(i)}\bigr)$.

\end{algorithmic}
\end{algorithm}

\subsubsection{First Step: Data Generation}
For the continuous-time system given in~\eqref{eq:continuous_nonlinear_system_2}, we perform a series of experiments to generate two datasets used for learning the system dynamics. In the first set of experiments, a constant initial condition is applied, and $(N+1)$ distinct control inputs, indexed by $q = 0, \dots, N$, are used. For each trial, the system states $x(0)$, $x(1)$, and $x(2)$ are recorded. This process is repeated for $M$ different initial conditions sampled from the domain $\cX$, indexed by $p = 1, \dots, M$.
The second dataset is constructed using the intermediate states $x(1)$ and the corresponding control inputs $u(1)$ obtained from the first dataset. These serve as the initial conditions and inputs for a new set of experiments in which only one time step of the system's response is recorded. This additional dataset is used to estimate the fractional order of the system by exploiting its memory-dependent behavior.

\subsubsection{Second Step: Dynamic Inference}
In this step, we begin by estimating the fractional order of the system by exploiting the memory-dependent property of the continuous-time fractional-order model described in~\eqref{eq:continuous_nonlinear_system_2}. Once the fractional order is identified, the control vector field is reconstructed using the input-output data collected in the first step. After the control vector field has been determined, the drift vector field is obtained using the known system responses and the previously estimated quantities.

\subsubsubsection{Fractional Order of the System}
Since only two time steps of the system dynamics is used to generate the first dataset, we simplify system~\eqref{eq:continuous_nonlinear_system_2} for 
$k=0$, $k=1$, and $k=2$ as follows:
\begin{align}
    x(1)^{(i,j)} & = h^{\alpha} \left( f(x(0)^{(i)}) + g(x(0)^{(i)}) u(0)^{(i,j)} \right) + x(0)^{(i)} \label{eq:continuous_nonlinear_system_3} \\
    x(2)^{(i,j)} & = x(0)^{(i)} + h^{\alpha} \left( f(x(1)^{(i,j)}) + g(x(1)^{(i,j)}) u(1)^{(i,j)} \right) + \psi(\alpha, 1)\left( x(0)^{(i)} - x(1)^{(i,j)} \right) \label{eq:continuous_nonlinear_system_4}
\end{align}
Applying the same procedure to the second dataset, we obtain:
\begin{align}\label{eq:order_inference_continuous_1}
    \Tilde{x}(2)^{(i,j)} = h^{\alpha} \left( f(x(1)^{(i,j)}) + g(x(1)^{(i,j)}) u(1)^{(i,j)} \right) + x(1)^{(i,j)}.
\end{align}
By subtracting~\eqref{eq:order_inference_continuous_1} from~\eqref{eq:continuous_nonlinear_system_4}, we obtain:
\begin{align}
    x(2)^{(i,j)} - \Tilde{x}(2)^{(i,j)} = \left( I_{n \times n} + \psi(\alpha, 1) \right) \left( x(0)^{(i)} - x(1)^{(i,j)} \right),
\end{align}
where $\psi(\alpha, 1) = \operatorname{diag}\left( \psi(\alpha_1,1), \psi(\alpha_2,1), \dots, \psi(\alpha_n,1) \right)$.\\
Thus, the values of $\alpha_i$ can be determined by solving the above equation.

\begin{remark}
The proposed method for estimating the system order also applies to integer -order systems, where the right-hand side of~\eqref{eq:order_inference_continuous_1} becomes zero when the order is one. This is because integer-order systems do not exhibit memory effects.
\end{remark}
\begin{remark}
While only four data points are theoretically sufficient to determine the system order under noiseless conditions, it is generally advisable to use longer time series to improve robustness against measurement noise.
\end{remark}

\subsubsubsection{Control Vector Field}\label{section:control-vector-field-continuous}
Let us define $y^{(i,j)}$ as 
\begin{align}\label{eq:y_continuous_1}
    y^{(i,j)} = h^{-\alpha} \left( x(1)^{(i,j)} - x(1)^{(i,0)} \right), \quad j = 1,\dots,N,\; i = 1,\dots,M.
\end{align}
By substituting $x(1)^{(i,j)}$ and $x(1)^{(i,0)}$ from the system dynamics into~\eqref{eq:y_continuous_1}, we obtain
\begin{align}\label{eq:y_continuous_2}
    y^{(i,j)} = g(x(0)^{(i)}) \left( u(0)^{(i,j)} - u(0)^{(i,0)} \right).
\end{align}
In the above equation, $g(\cdot) \in \mathbb{R}^{n \times m}$ is an unknown matrix-valued function. We approximate $g(\cdot)$ using a truncated series expansion of orthonormal basis functions. The corresponding coefficients are estimated by solving a least-squares optimization problem.
To do so, we excite the system using only one control input $u_{\ell}(0)$ at a time, setting all others to zero. This isolates the $\ell$-th column of $g(\cdot)$ in each iteration. Thus, for each $\ell = 1, \dots, m$, we denote $u_{\ell}(k) \in \mathbb{R}$ and $g_{\ell}(\cdot) = [g_{1,\ell}(\cdot), \dots, g_{n,\ell}(\cdot)]^\intercal$.
Each component $g_{i,\ell}$ is approximated by a truncated orthonormal series~\cite{akccay1999orthonormal}:
\begin{align}\label{eq:g_approx}
    \hat{g}_{i,\ell}(x) \approx \sum_{l = 1}^{L} \beta_{i,\ell}^{(l)} \phi_{i,\ell}^{(l)}(x), \quad i = 1,\dots,n,
\end{align}
where $\{\phi_{i,\ell}^{(l)}(\cdot)\}_{l=1}^{L}$ are orthonormal basis functions and $\beta_{i,\ell}^{(l)}$ are unknown coefficients.
Substituting~\eqref{eq:g_approx} into~\eqref{eq:y_continuous_2} gives:
\begin{align}\label{eq:y_continuous_3}
    y_{\ell}^{(i,j)} = 
    \begin{bmatrix}
        \sum_{l = 1}^{L} \beta_{1,\ell}^{(l)} \phi_{1,\ell}^{(l)}(x(0)^{(i)}) \\
        \vdots \\
        \sum_{l = 1}^{L} \beta_{n,\ell}^{(l)} \phi_{n,\ell}^{(l)}(x(0)^{(i)})
    \end{bmatrix} 
    \Delta u_{\ell}(0)^{(i,j)},
\end{align}
where $\Delta u_{\ell}(0)^{(i,j)} = u_{\ell}(0)^{(i,j)} - u_{\ell}(0)^{(i,0)}$.
Since the basis functions are orthonormal, the unknowns are the coefficients $\beta_{i,\ell}^{(l)}$. Define:
\begin{align*}
    \phi_{\ell}^{(i,j)} & = 
    \begin{bmatrix}
        \phi_{1,\ell}(x(0)^{(i)}) & \mathbf{0} & \cdots & \mathbf{0} \\
        \mathbf{0} & \phi_{2,\ell}(x(0)^{(i)}) & \cdots & \mathbf{0} \\
        \vdots & \vdots & \ddots & \vdots \\
        \mathbf{0} & \mathbf{0} & \cdots & \phi_{n,\ell}(x(0)^{(i)})
    \end{bmatrix} 
    \Delta u_{\ell}(0)^{(i,j)} \in \mathbb{R}^{n \times (nL)}, \\
    \beta_{r,\ell} & = \begin{bmatrix} \beta_{r,\ell}^{(1)} & \cdots & \beta_{r,\ell}^{(L)} \end{bmatrix}^\intercal \in \mathbb{R}^{L}, \quad r = 1, \dots, n.
\end{align*}
Stacking all samples, we define:
\begin{align}\label{eq:26}
    \Phi_{\ell} &= 
    \begin{bmatrix}
        \phi_{\ell}^{(1,1)} \\
        \vdots \\
        \phi_{\ell}^{(M,N)}
    \end{bmatrix} \in \mathbb{R}^{(MNn) \times (nL)}, \quad
    Y_{\ell} = 
    \begin{bmatrix}
        y_{\ell}^{(1,1)} \\
        \vdots \\
        y_{\ell}^{(M,N)}
    \end{bmatrix} \in \mathbb{R}^{(MNn)}, \\
    B_{\ell} &=
    \begin{bmatrix}
        \beta_{1,\ell} \\
        \vdots \\
        \beta_{n,\ell}
    \end{bmatrix} \in \mathbb{R}^{(nL)}.
\end{align}
The regression problem is formulated as:
\begin{align}
    Y_{\ell} = \Phi_{\ell} B_{\ell}.
\end{align}
The coefficients are estimated by solving the least-squares problem:
\begin{align}\label{eq:optimization_problem}
    B_{\ell}^* = \underset{B_{\ell}}{\arg\min} \; \|Y_{\ell} - \Phi_{\ell} B_{\ell}\|^2.
\end{align}

The closed-form solution is:
\begin{align}
    B_{\ell}^* = (\Phi_{\ell}^\intercal \Phi_{\ell})^{-1} \Phi_{\ell}^\intercal Y_{\ell}.
\end{align}

\subsubsubsection{Drift Vector Field}\label{section:drift-vector-field-continuous}
To determine the drift-vector field, we use equation~\eqref{eq:continuous_nonlinear_system_3}. Since both the control-vector field $g(\cdot)$ and the system order $\alpha$ have already been estimated, the only remaining unknown term in this equation is the drift function $f(x(0))$. It can be directly computed by rearranging~\eqref{eq:continuous_nonlinear_system_3} as follows:
\begin{align}
    f(x(0)^{(i)}) = h^{-\alpha} \left( x(1)^{(i,j)} - x(0)^{(i)} \right) - g(x(0)^{(i)}) u(0)^{(i,j)}.
\end{align}
Since $x(0)^{(i)}$ has been sampled uniformly from the domain $\cX$, this procedure ensures that the reconstructed function $f(\cdot)$ is well-defined and valid over the entire state space.

\begin{remark}
The optimization problem in~\eqref{eq:optimization_problem} used to estimate the control-vector field is a linear regression problem with a convex quadratic cost function. Therefore, the solution is guaranteed to be globally optimal. As a result, the identified functions $g(\cdot)$ and $f(\cdot)$ are unique, and the overall system dynamics can be reconstructed with guaranteed consistency.
\end{remark}

\subsection{LDF: Learning Discrete-time Fractional-Order Dynamics}
Algorithm~\ref{algorithm:discrete-systems} presents the Learning Discrete-time Fractional-order dynamics (LDF) method for inferring the dynamics of discrete-time fractional-order nonlinear systems. The LDF framework follows the same two-step procedure as its continuous-time counterpart: data generation through input-output experiments and dynamic inference via linear regression to estimate the system's fractional order, drift vector field, and control vector field.

\subsubsection{First Step: Data Generation}
For the discrete-time system in~\eqref{eq:discrete_nonlinear_system_2}, we design experiments to generate three datasets that will be used to infer the system dynamics. We begin by selecting a fixed initial condition and applying $N + 1$ different control inputs, indexed by $q = 0, \dots, N$. For each input, we record the corresponding states $x(0)$, $x(1)$, and $x(2)$. This experiment is repeated for $M$ different initial conditions sampled from the set $\cX$, indexed by $p = 1, \dots, M$.
The second dataset is generated by using the states $x(1)$ and control inputs $u(1)$ from the first dataset as the new initial conditions and control inputs, respectively. In this case, only one step of the system response is recorded, which is used to estimate the system's fractional order.
Finally, the third dataset is generated by setting the initial condition to $x(2)$ and the control input to $u(2)$, again recording only a single step of the response. This final dataset further reinforces the estimation of the system’s memory-dependent behavior.

\subsubsection{Second Step: Dynamic Inference}
\subsubsubsection{Fractional Order of the System}
Since we only use two time steps of the system dynamics to generate the first dataset, we simplify equation~\eqref{eq:general_discrete_nonlinear_system_2} for $k = 0$, $k = 1$, and $k = 2$

\begin{algorithm}[!h]
\caption{Learning Discrete-time Fractional-order dynamics (LDF)}
\label{algorithm:discrete-systems}
\begin{algorithmic}[1]
\STATE \textbf{Input:}
\begin{itemize}[leftmargin=*]
  \item $\{x(0)^{(i)}\}_{i=1}^{M}$: A set of $M$ initial conditions uniformly sampled from $\cX \subseteq \mathbb{R}^n$.
  \item $\{U(0)^{(i,j)}\}$, $\{U(1)^{(i,j)}\}$, and $\{U(2)^{(i,j)}\}$: three sets of randomly chosen system inputs, used to excite the system over multiple trials.
  \item $\phi_i^{(1)}, \ldots, \phi_i^{(L)}$, $i = 1, \ldots, n$: A collection of orthonormal basis functions with $L$ terms in the truncated series to approximate the unknown functions.
\end{itemize}
\STATE \textbf{Output:} Functions $f(\cdot)$, $g(\cdot)$, and $\alpha$.

\FOR{$\ell = 1, \ldots, m$}
    \STATE \textbf{Set inputs for the $\ell$-th element only:}
    For each $i,j$, set 
    $u_{\ell}(0)^{(i,j)} = U_{\ell}(0)^{(i,j)}$, 
    $u_{\ell}(1)^{(i,j)} = U_{\ell}(1)^{(i,j)}$, 
    $u_{\ell}(2)^{(i,j)} = U_{\ell}(2)^{(i,j)}$.
    For all other elements $q \neq \ell$, set 
    $u_{q}(0)^{(i,j)} = u_{q}(1)^{(i,j)} = u_{q}(2)^{(i,j)} = 0.$
    \STATE \textbf{Data Generation:} 
    Run the system for three consecutive time steps. 
    At $k=0\to1$, start from $x(0)^{(i)}$ and apply $u(0)^{(i,j)}$; record $x(1)^{(i,j)}$. 
    At $k=1$, apply $u(1)^{(i,j)}$; record $x(2)^{(i,j)}$. 
    At $k=2$, apply $u(2)^{(i,j)}$; record $x(3)^{(i,j)}$.
    \STATE \textbf{System's Order Determination:} 
    Reinitialize the system at $x(1)^{(i,j)}$. 
    Use $u(1)^{(i,j)}$ for one step to get $\Tilde{x}(2)^{(i,j)}$, 
    then apply $u(2)^{(i,j)}$ for the next step to obtain $\Tilde{x}(3)^{(i,j)}$.
    \IF{$\ell = 1$}
        \STATE Determine the system's order $\alpha$ of the system by solving the following equation:
        \[
        x(2)^{(i,j)} - \Tilde{x}(2)^{(i,j)} = \frac{1}{2}(A - A^2)x(0)^{(i)},
        \]
        \[
        x(3)^{(i,j)} - \Tilde{x}(3)^{(i,j)} = \frac{1}{2}(A - A^2)x(1)^{(i,j)} + \frac{1}{6}(A^3 - 3A^2 + 2A)x(0)^{(i)}.
        \]
    \ENDIF
    \STATE Reconstruction of the control-vector and drift-vector fields:
    \begin{itemize}
    \item 
    
    Compute $y^{(i,j)} = x(1)^{(i,j)} - x(1)^{(i,0)}$ for $i = 1, \ldots, M$, $j = 1, \ldots, N$.
    \item Form the matrix $Y = [y^{(1,1)}, \ldots, y^{(M,1)}, \ldots, y^{(1,N)}, \ldots, y^{(M,N)}]^\intercal$.
    \item Define $\phi^{(i,j)}$ as:
    \[
    \phi^{(i,j)} =
    \begin{bmatrix}
        \phi_1(x(0)^{(i)}) & \mathbf{0}_{1 \times L} & \cdots & \mathbf{0}_{1 \times L} \\
        \mathbf{0}_{1 \times L} & \phi_2(x(0)^{(i)}) & \cdots & \mathbf{0}_{1 \times L} \\
        \vdots & \vdots & \ddots & \vdots \\
        \mathbf{0}_{1 \times L} & \mathbf{0}_{1 \times L} & \cdots & \phi_n(x(0)^{(i)})
    \end{bmatrix}
    \Delta u(0)^{(i,j)},
    \]
    where $\Delta u(0)^{(i,j)} = u(0)^{(i,j)} - u(0)^{(i,0)}$.
    \item Form the matrix $\Phi = [\phi^{(1,1)}, \ldots, \phi^{(M,1)}, \ldots, \phi^{(1,N)}, \ldots, \phi^{(M,N)}]^\intercal$.
    \item Solve for $B = (\Phi^\intercal \Phi)^{-1} \Phi^\intercal Y$.
    \item Define $g_{\ell} = [\beta_1^\intercal \phi_1, \ldots, \beta_n^\intercal \phi_n]^\intercal$.
    \end{itemize}
\ENDFOR
\STATE Return $g = [g_1, \ldots, g_m]$ and $f(x(0)^{(i)}) = -g(x(0)^{(i)})u(0)^{(i,1)} + h^{-\alpha}(x(1)^{(i,1)} - x(0)^{(i)})$.
\end{algorithmic}
\end{algorithm}

\clearpage
\noindent as follows:
\begin{align}
    x(1)^{(i,j)} & = f(x(0)^{(i)}) + g(x(0)^{(i)})u(0)^{(i,j)} - \psi(\alpha,1) x(0)^{(i)} \nonumber \\
    & = f(x(0)^{(i)}) + g(x(0)^{(i)})u_0^{(i,j)} + A x(0)^{(i)}, \label{eq:nonlin_disc_sys_3} \\
    x(2)^{(i,j)} & = f(x(1)^{(i,j)}) + g(x(1)^{(i,j)})u(1)^{(i,j)} - \psi(\alpha,1) x(1)^{(i,j)} - \psi(\alpha,2) x(0)^{(i)} \nonumber \\
    & = f(x(1)^{(i,j)}) + g(x(1)^{(i,j)})u(1)^{(i,j)} + A x(1)^{(i,j)} + \tfrac{1}{2}(A - A^2)x(0)^{(i)}, \label{eq:nonlin_disc_sys_4} \\
    x(3)^{(i,j)} & = f(x(2)^{(i,j)}) + g(x(2)^{(i,j)})u(2)^{(i,j)} - \psi(\alpha,1) x(2)^{(i,j)} - \psi(\alpha,2) x(1)^{(i,j)} - \psi(\alpha,3) x(0)^{(i)} \nonumber \\
    & = f(x(2)^{(i,j)}) + g(x(2)^{(i,j)})u(2)^{(i,j)} + A x(2)^{(i,j)} + \tfrac{1}{2}(A - A^2)x(1)^{(i,j)} \nonumber\\
    & \quad + \tfrac{1}{6}(A^3 - 3A^2 + 2A)x(0)^{(i)}, \label{eq:nonlin_disc_sys_5}
\end{align}
where $A \in \mathbb{R}^{n \times n}$ is a diagonal matrix with $\alpha_i$ on its principal diagonal:
\begin{align}
    A = \operatorname{diag}(\alpha_1, \alpha_2, \dots, \alpha_n).
\end{align}
Applying the same procedure to the second and third datasets, we obtain:
\begin{align}
    \Tilde{x}(2)^{(i,j)} & = f(x(1)^{(i,j)}) + g(x(1)^{(i,j)})u(1)^{(i,j)} - \psi(\alpha,1) x(1)^{(i,j)} \nonumber \\
    & = f(x(1)^{(i,j)}) + g(x(1)^{(i,j)})u(1)^{(i,j)} + A x(1)^{(i,j)}, \label{eq:nonlin_disc_sys_6} \\
    \Tilde{x}(3)^{(i,j)} & = f(x(2)^{(i,j)}) + g(x(2)^{(i,j)})u(2)^{(i,j)} - \psi(\alpha,1) x(2)^{(i,j)} \nonumber \\
    & = f(x(2)^{(i,j)}) + g(x(2)^{(i,j)})u(2)^{(i,j)} + A x(2)^{(i,j)}. \label{eq:nonlin_disc_sys_7}
\end{align}
Subtracting equations~\eqref{eq:nonlin_disc_sys_6} and~\eqref{eq:nonlin_disc_sys_7} from~\eqref{eq:nonlin_disc_sys_4} and~\eqref{eq:nonlin_disc_sys_5}, respectively, yields:
\begin{align}
    x(2)^{(i,j)} - \Tilde{x}(2)^{(i,j)} & = \tfrac{1}{2}(A - A^2)x(0)^{(i)}, \label{eq:final_order_inference_1}\\
    x(3)^{(i,j)} - \Tilde{x}(3)^{(i,j)} & = \tfrac{1}{2}(A - A^2)x(1)^{(i,j)} + \tfrac{1}{6}(A^3 - 3A^2 + 2A)x(0)^{(i)}. \label{eq:final_order_inference_2}
\end{align}
In equation~\eqref{eq:final_order_inference_1}, both $x(2)^{(i,j)} - \Tilde{x}(2)^{(i,j)}$ and $x(0)^{(i)}$ are known. Therefore, the system's fractional order can be determined by solving this algebraic equation. Since the equation is quadratic in $\alpha_i$, it yields two solutions. We select the solution that also satisfies equation~\eqref{eq:final_order_inference_2}.

\subsubsubsection{Control Vector Field}
Following the same procedure outlined in Section~\ref{section:control-vector-field-continuous}, we define
\begin{align}\label{eq:y_discrete_1}
    y^{(i,j)} = x(1)^{(i,j)} - x(1)^{(i,0)}, \quad j = 1,\dots,N,\ i = 1,\dots,M,
\end{align}
and reconstruct the functions $g_{i,j}(\cdot)$ by expressing $y^{(i,j)}$ as a linear function of the perturbed inputs. As in the continuous-time case, orthonormal basis functions are employed to approximate $g_{i,j}(\cdot)$, and the unknown coefficients are estimated via least-squares regression.

\subsubsubsection{Drift Vector Field}
The drift functions $f_i(\cdot)$ can be reconstructed using the approach described in Section~\ref{section:drift-vector-field-continuous}, by applying equation~\eqref{eq:nonlin_disc_sys_3} and subtracting the known terms from both sides. Since the fractional order $\alpha$ and control-vector field $g(\cdot)$ have already been identified, the remaining unknown component in the system equation is $f(x(0))$, which can be directly computed.

\section{Experimental Results}\label{section:simulation}
In this section, the proposed algorithms are applied to fractio- nal-order nonlinear continuous-time and discrete-time systems to verify their effectiveness through simulation results. Additionally, an algorithm for integer-order systems from the literature is used to infer an integer-order model from a dataset generated by a fractional-order system. The responses of the identified integer-order and fractional-order systems are then compared to demonstrate the advantages of fractional-order models in capturing long-range dependent processes.
Simulations with noisy measurements are also included to evaluate the robustness of the proposed algorithm against measurement noise, as described in Remark~\ref{rem:noise} below.

\begin{remark}\label{rem:noise}
In all simulations involving measurement noise, the noisy state measurement at time $t$ (or step $k$) is modeled as $x_i(t) + 0.05 A_i v_i(t)$ for $i = 1, \ldots, n$, where $x_i(t)$ is the true state, $A_i$ is the amplitude of $x_i(t)$ at time $t$, and $v_i(t) \sim \mathcal{N}(0, 1)$ is additive white Gaussian noise with zero mean and unit variance.
\end{remark}

\subsection{Numerical Simulations for Learning the Dynamics of Fractional-Order Systems}

\subsubsection{Van der Pol Oscillator}\label{section:simulation_vanderpol_system}
\begin{figure}[!h]
\centering
\includegraphics[width=0.48\columnwidth]{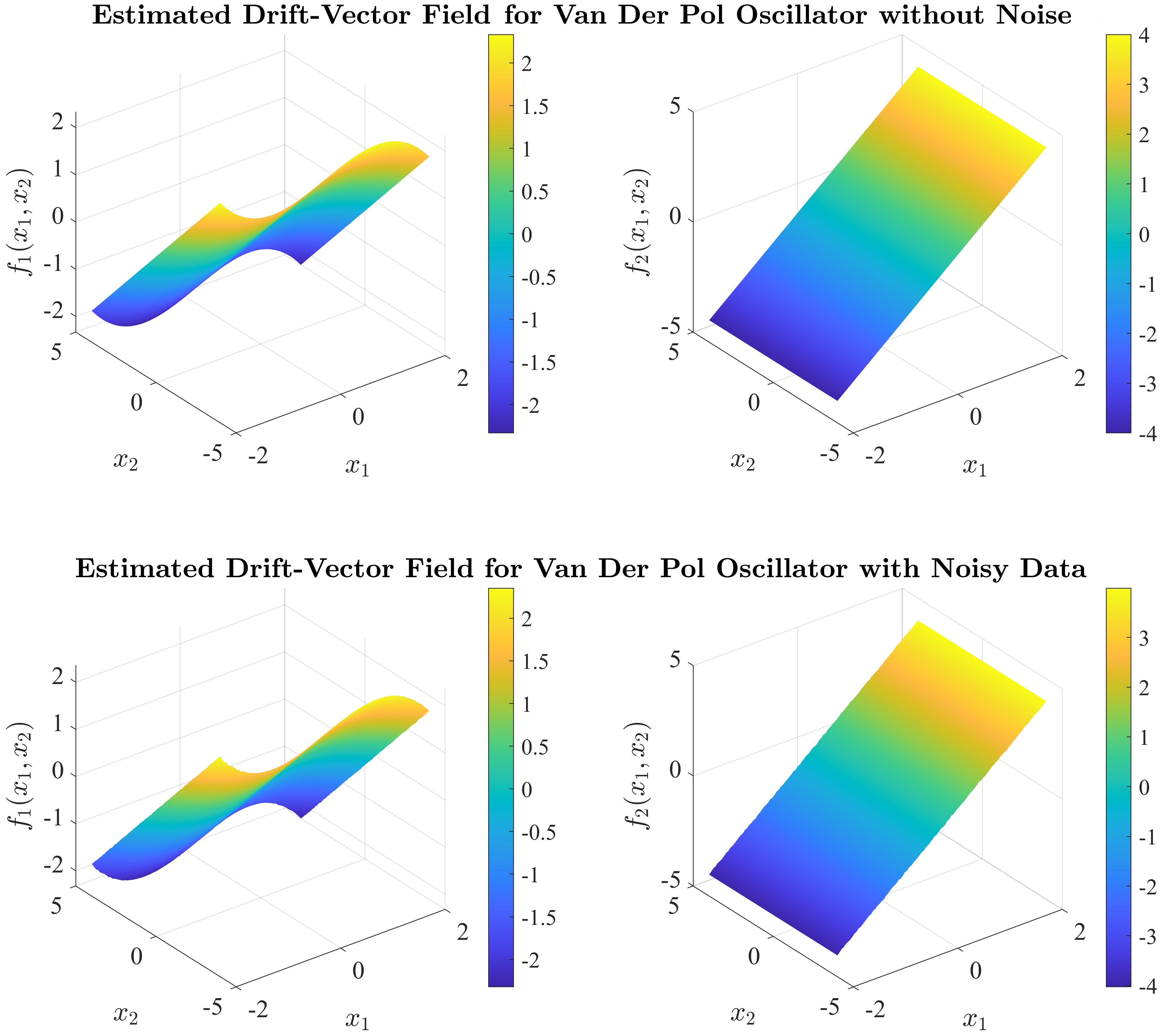} \hfill
\includegraphics[width=0.48\columnwidth]{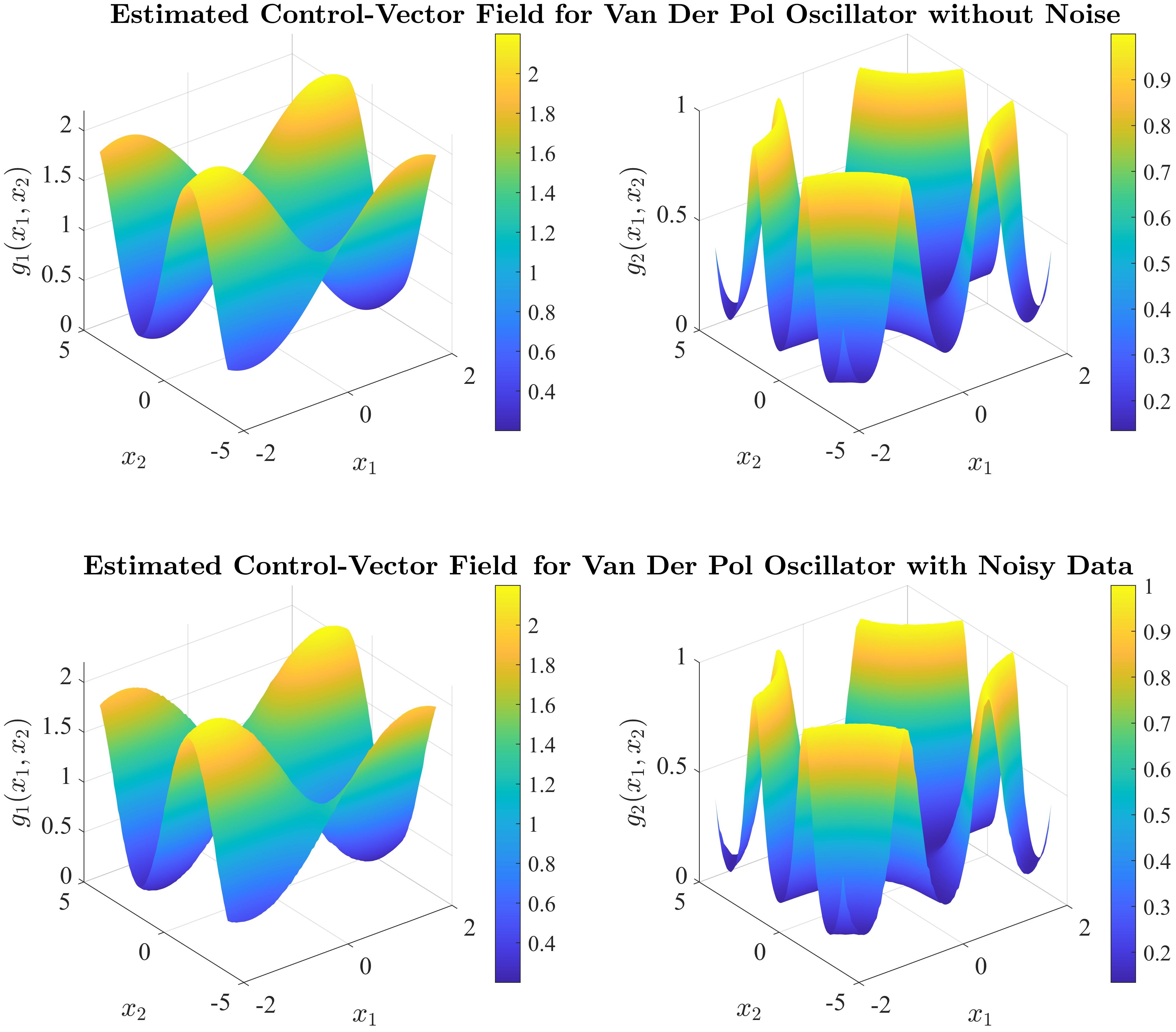}
\caption{Estimated Drift-Vector and Control-Vector Fields for Van der Pol oscillator.}
\label{fig:vanderpol_DVF}
\end{figure}

The Van der Pol oscillator is a non-conservative oscillatory system with nonlinear damping. Its fractional-order variant can be expressed as follows:
\begin{align}\label{eq:vanderpol_oscillator}
\begin{cases}
D_t^{\alpha}x_1 = \epsilon \left(x_1 - \frac{1}{3}x_1^3 - x_2\right) + \left(1 + \exp(\sin(x_1))\right) u(t), \\
D_t^{\alpha}x_2 = \frac{1}{\epsilon} x_1 + \exp\left(\sin(x_2 - 0.5\pi) - 1\right) u(t).
\end{cases}
\end{align}
The system parameters are set to $\epsilon = 0.5$ and $\alpha = 0.9$. Using system~\eqref{eq:vanderpol_oscillator}, we generate the datasets required for dynamic inference over the domain $x \in [-2, 2] \times [-4, 4]$. The drift vector and control vector fields are defined as follows:
\begin{align*}
    f(x) & = \begin{bmatrix}
        0.5\left(x_1 - \frac{1}{3}x_1^3 - x_2\right) \\
        2x_1
    \end{bmatrix}, \\
    g(x) & = \begin{bmatrix}
        1.2 + \sin(x_1)\sin(x_2) \\
        \exp\left(\sin(x_2 - 0.5\pi) - 1\right)
    \end{bmatrix}.
\end{align*}
In these simulations, the number of expansion terms is set to $L = 5$. The numerical simulation results are presented in Figure~\ref{fig:vanderpol_DVF}. Simulations with noisy measurements are also included to evaluate the robustness of the proposed algorithm to measurement noise. Additionally, the approximation errors for the drift and control vector fields of the Van der Pol oscillator are provided in Appendix~\ref{section:experiments_cont}.

\subsubsection{Lotka-Volterra Model}\label{section:simulation_lotka_system}
The Lotka-Volterra equations describe an ecological pred- ator-prey model. This system can be modeled using the following dynamical equations:
\begin{align}\label{eq:lotka-volterra}
\begin{cases}
D_t^{\alpha}x_1 = a x_1 - \beta x_1 x_2 + \left(4 + \sin(x_1)\exp(-1 + \cos(x_2))\right) u(t), \\
D_t^{\alpha}x_2 = \delta x_1 x_2 - \gamma x_2 + \left(1 + \exp(\sin(x_1)\cos(x_2))\right) u(t).
\end{cases}
\end{align}
We set the system parameters to $a = \beta = 0.5$, $\delta = 1.3$, and $\gamma = 0.6$. System~\eqref{eq:lotka-volterra} is used to generate the datasets required for dynamic inference over the domain $x \in [-2, 2] \times [-4, 4]$. The drift vector and control vector fields are given by:
\begin{align*}
    f(x) & = \begin{bmatrix}
        0.5 x_1 - 0.5 x_1 x_2 \\
        1.3 x_1 x_2 - 0.6 x_2
    \end{bmatrix}, \\
    g(x) & = \begin{bmatrix}
        4 + \sin(x_1)\exp(-1 + \cos(x_2)) \\
        1 + \exp(\sin(x_1)\cos(x_2))
    \end{bmatrix}.
\end{align*}
In these simulations, the number of expansion terms in the truncated series is set to $L = 5$. The numerical results are presented in Figure~\ref{fig:vanderpol_DVF}. Simulations with noisy measurements are also included to assess the robustness of the algorithm against measurement noise.
The estimation errors for the drift vector and control vector fields of the Lotka-Volterra model are provided in Appendix~\ref{section:experiments_cont}.

\begin{figure}[!h]
\centering
\includegraphics[width=0.48\columnwidth]{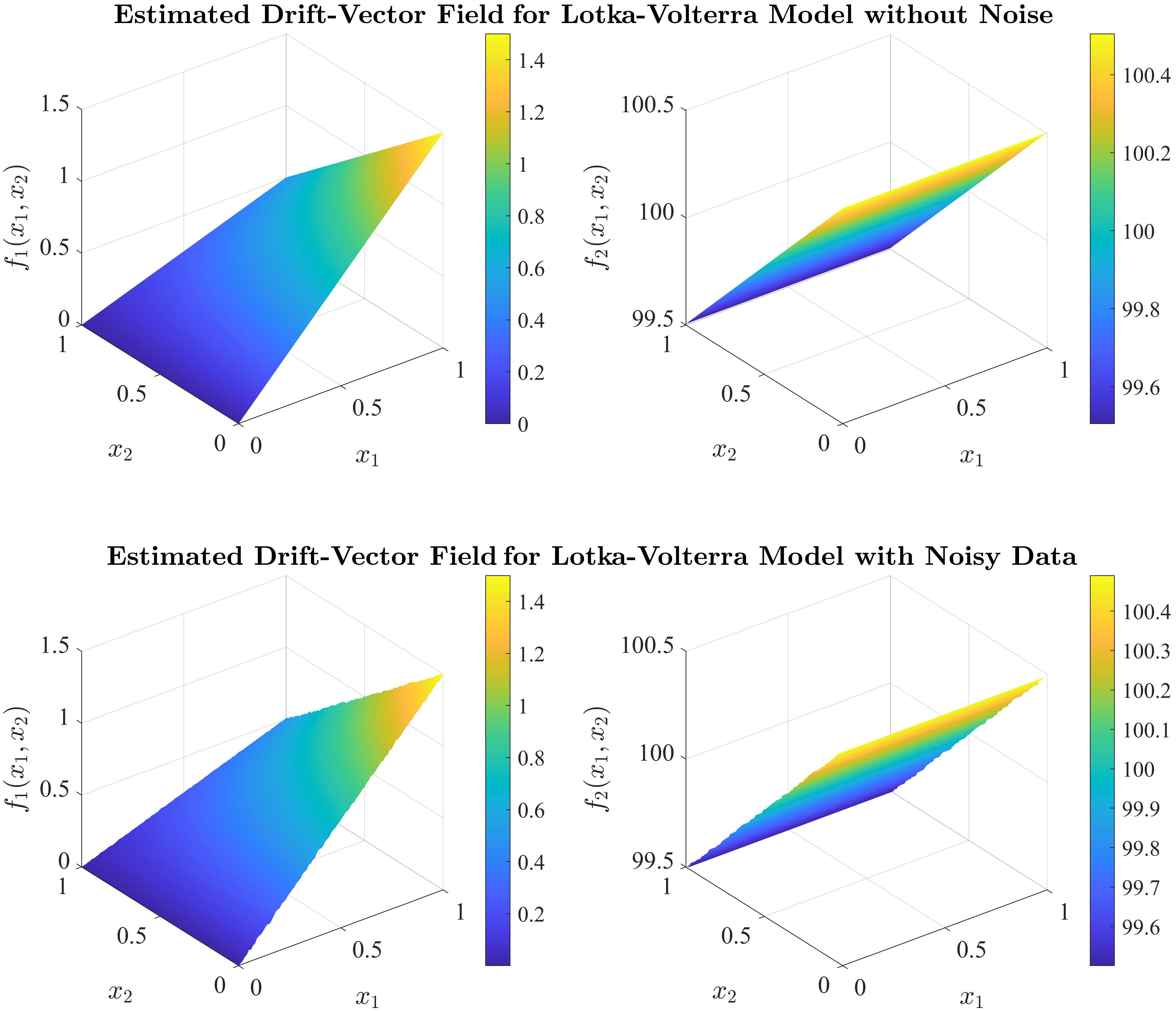} \hfill
\includegraphics[width=0.48\columnwidth]{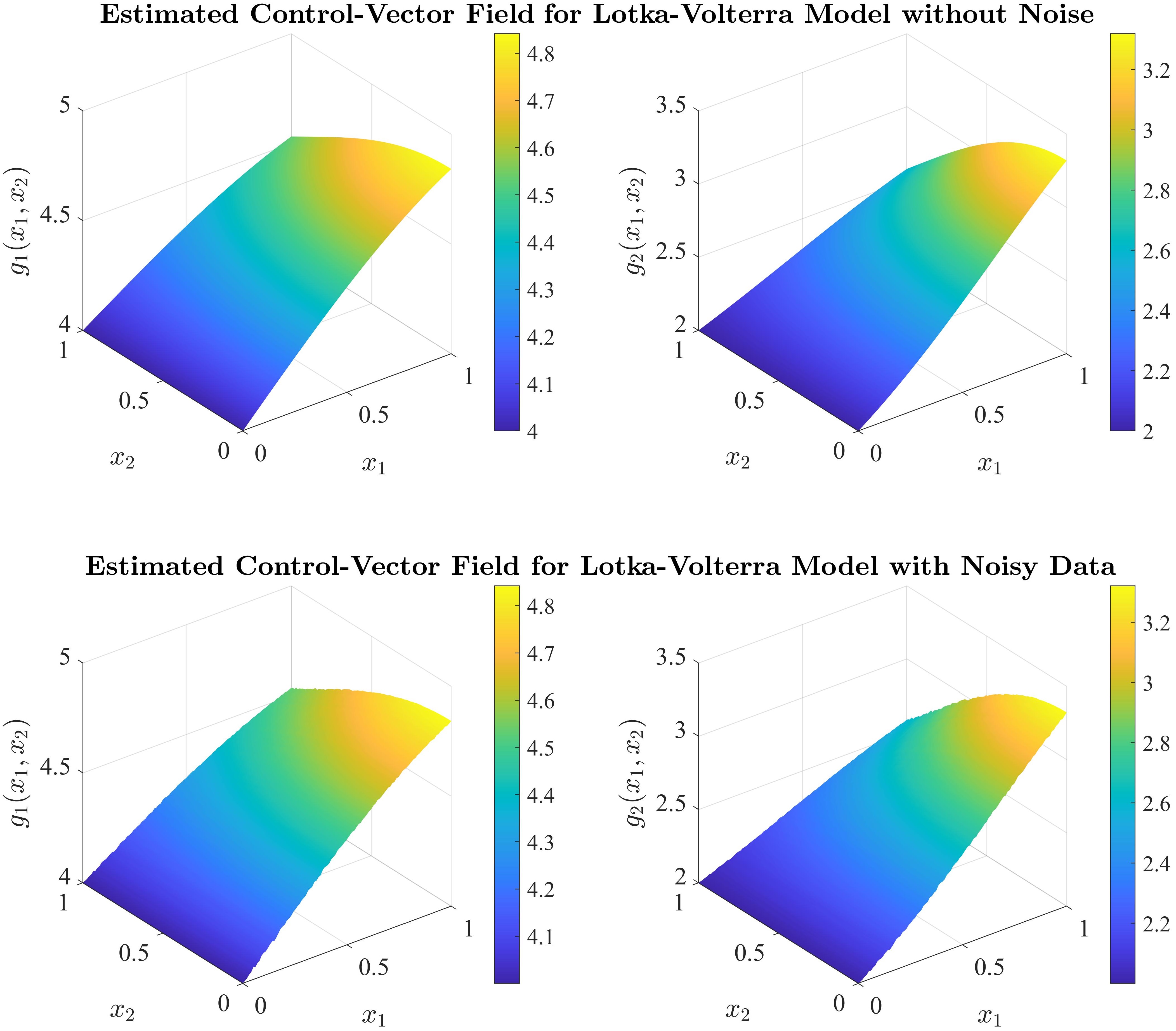}
\caption{Estimated Drift-Vector and Control-Vector Fields for Lotka-Volterra Model.}
\label{fig:lotka_DVF}
\end{figure}

\subsubsection{Logistic Map}\label{section:simulation_logistic_map}
Consider the following discrete-time fractional-order logistic map:
\begin{align}\label{eq:logistic_map}
    \Delta^{\alpha} x(k+1) = \mu x(k)(1 - x(k)) + \left(1 - \cos(x(k)) \exp\left(3(\sin(x(k) - 0.7\pi) - 1)\right)\right) u(k).
\end{align}
The classical (integer-order) logistic map was originally introduced in the late 1940s for pseudo-random number generation. Its fractional-order counterpart has since received considerable attention in the literature~\cite{munkhammar2013chaos}. Given the structure of equation~\eqref{eq:26} and the optimization problem~\eqref{eq:optimization_problem}, the scalar system in~\eqref{eq:logistic_map} provides a suitable testbed for validating our theoretical results.

\begin{figure}[!h]
\centering
\includegraphics[width=0.48\columnwidth]{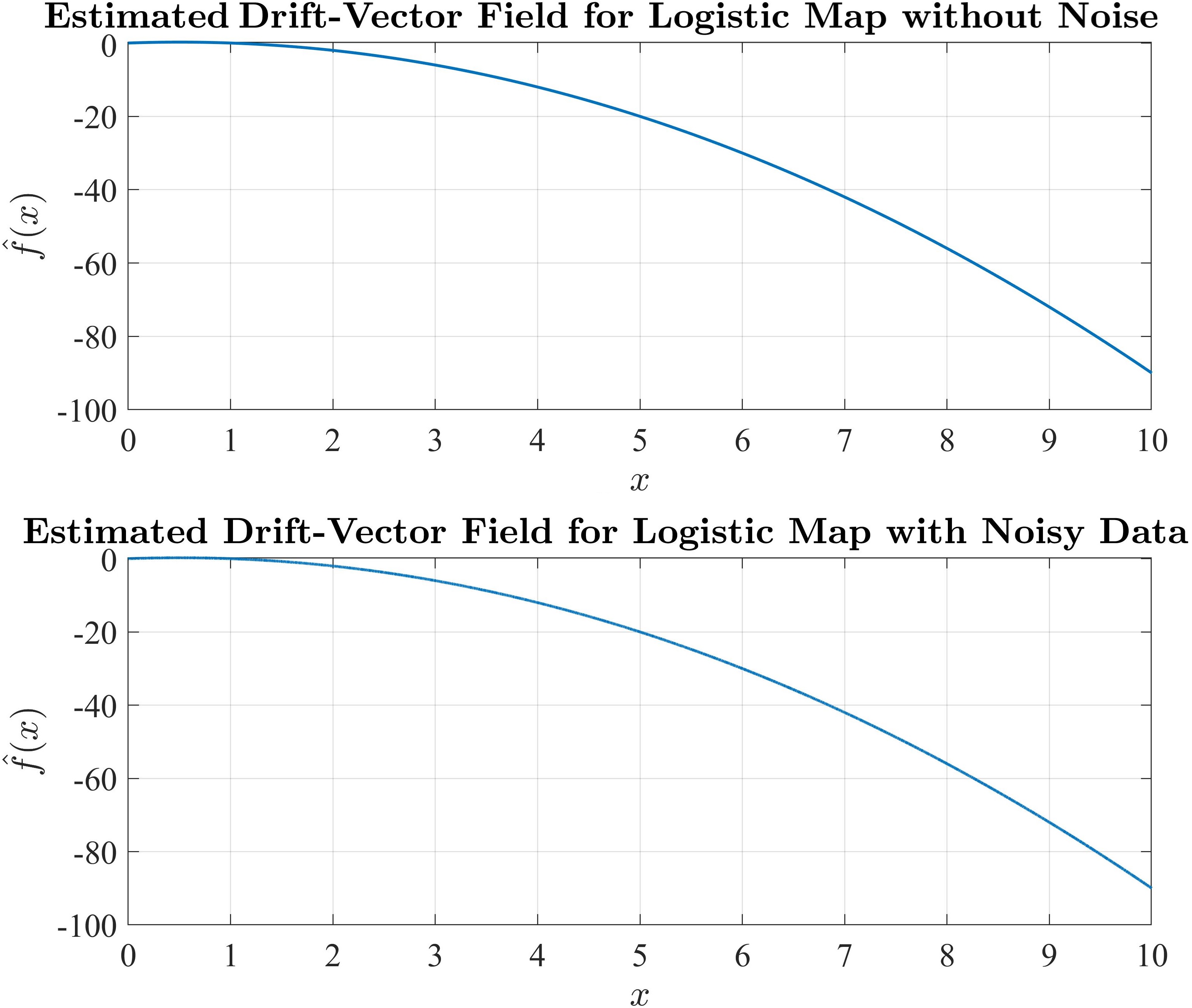} \hfill
\includegraphics[width=0.48\columnwidth]{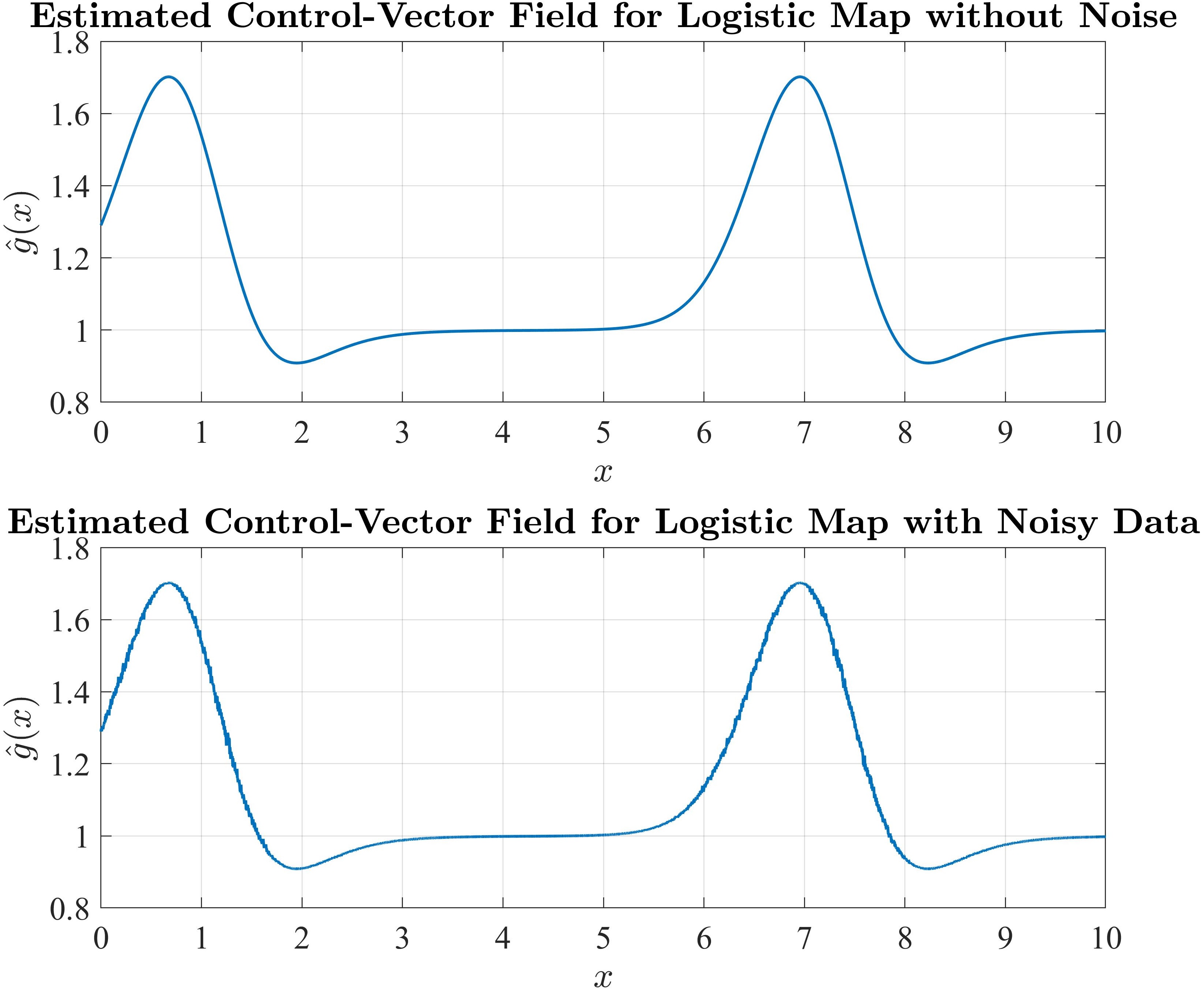}
\caption{Estimated Drift-Vector and Control-Vector Fields for Logistic Map.}
\label{fig:logistic_DVF}
\end{figure}

We use system~\eqref{eq:logistic_map} to generate the datasets required for dynamic inference over the domain $x \in [0, 8]$, with the parameters $\alpha = 0.6$, $\mu = 1$, $a = -1$, $b = 4$, and $c = 0.7$. The drift vector and control vector fields are given by:
\begin{align*}
    f(x(k)) & = x(k)(1 - x(k)), \\
    g(x(k)) & = 1 - \cos(x(k)) \exp\left(3(\sin(x(k) - 0.7\pi) - 1)\right).
\end{align*}
As described in Section~\ref{section:proposed_algorithm}, we approximate the control vector field using a truncated series of orthonormal basis functions. In these simulations, we use $L = 7$ expansion terms.
Simulation results for the drift vector field in both fractional-order and integer-order systems are presented in Figure~\ref{fig:logistic_DVF}. As shown, the proposed algorithm provides an accurate approximation of the true function. The maximum absolute error values for the estimated drift vector fields in the fractional-order and integer-order cases are $0.0038$ and $1.2566$, respectively.
This high-accuracy approximation is primarily attributed to two factors: (1) the large number of data points collected across the domain, and (2) the use of $7$ expansion terms in the truncated series.
To assess robustness against measurement noise, we also approximate the drift vector field using noisy observations. The corresponding simulation results are included in Figure~\ref{fig:logistic_DVF}.

\subsubsection{Ultra-Capacitor}\label{section:simulation_ultra_capacitor}
\begin{figure}[!h]
\centering
\includegraphics[width=0.48\columnwidth]{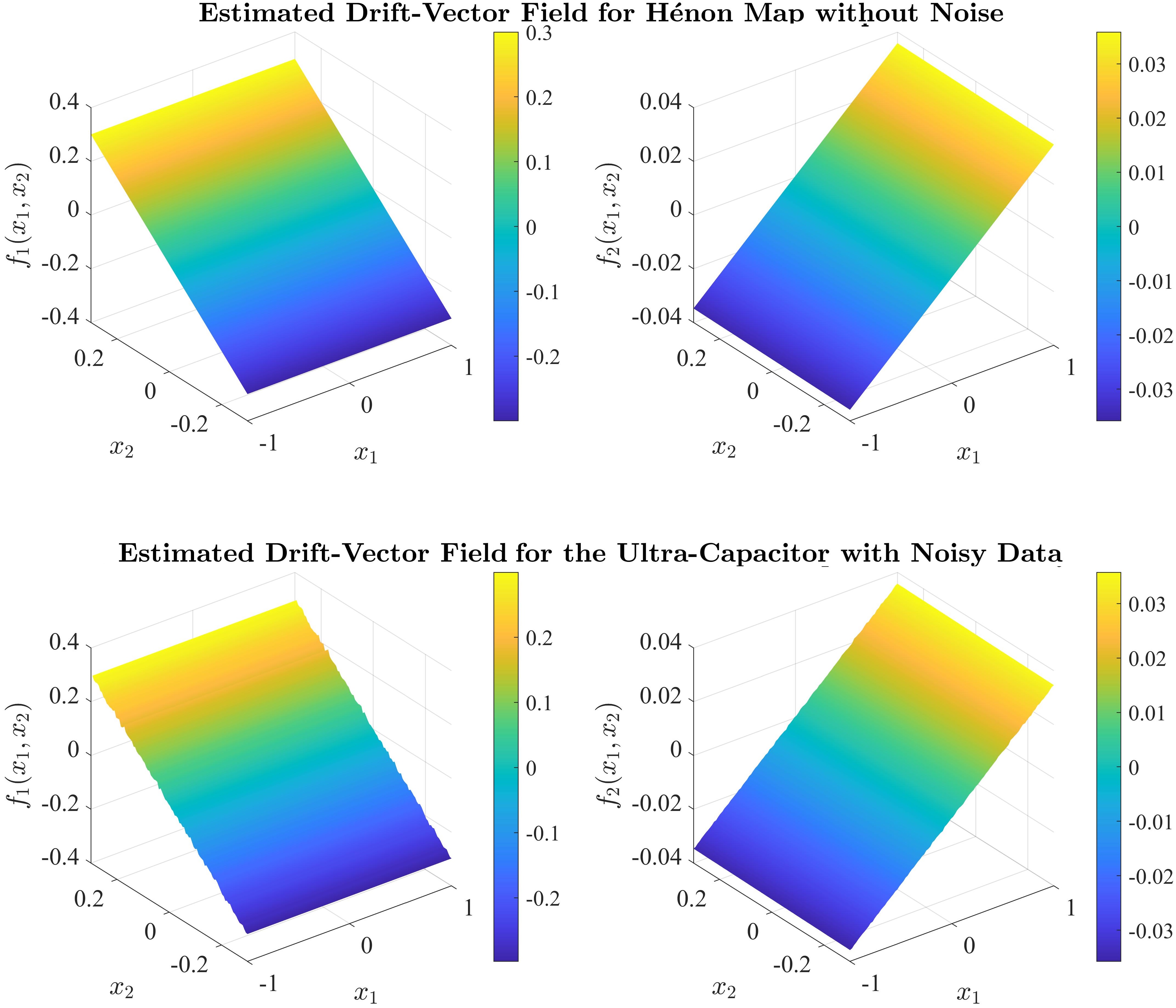} \hfill
\includegraphics[width=0.48\columnwidth]{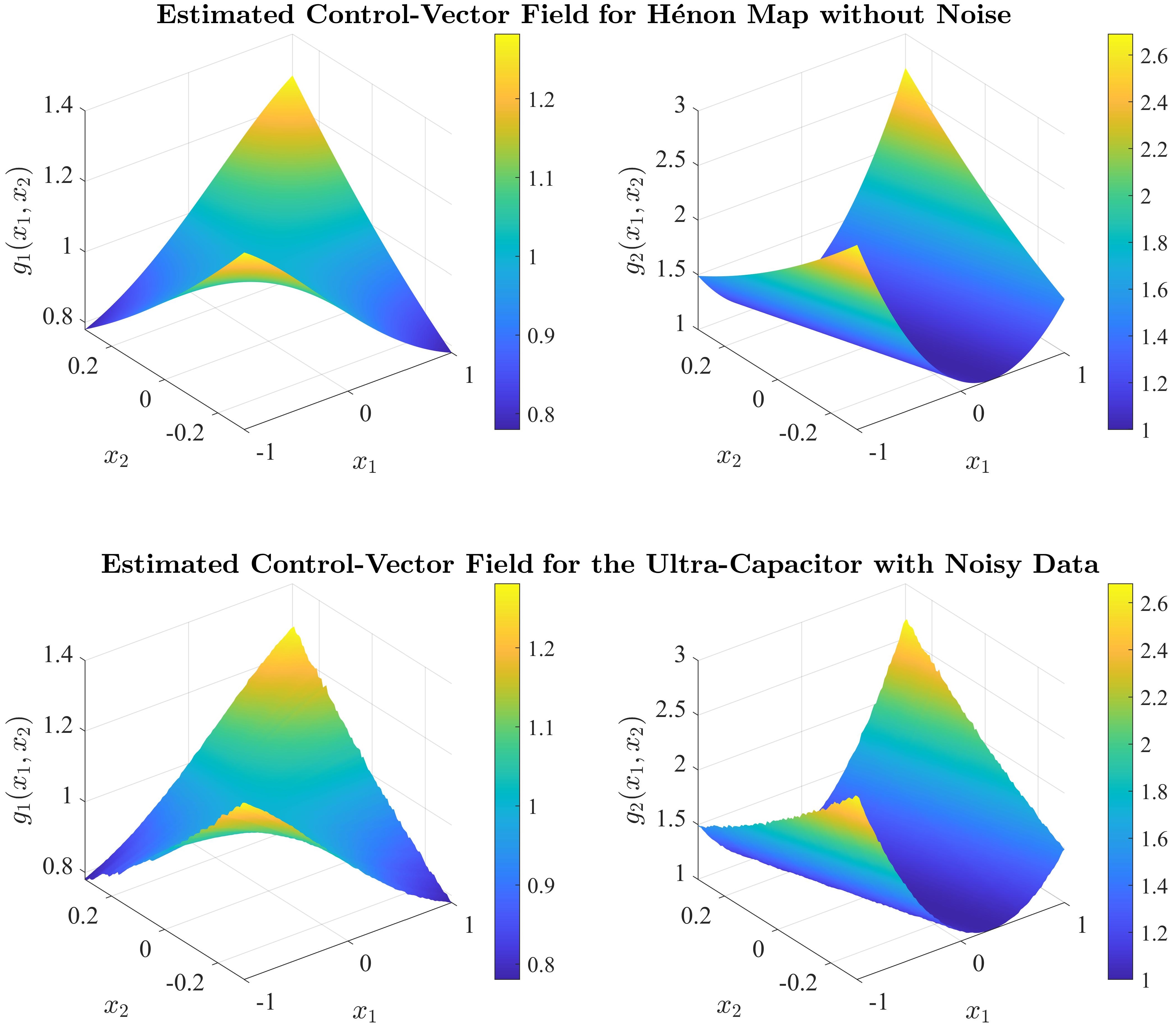}
\caption{Estimated Drift-Vector and Control-Vector Fields for the Ultra-Capacitor.}
\label{fig:capacitor_DVF}
\end{figure}

Consider the following fractional-order Ultra-Capacitor system \cite{sadeghian2013general}:
\begin{align}\label{eq:ultra_capacitor}
\left\{
\begin{array}{ll}
    \Delta^{\alpha} x_{1}(k+1) & = x_{2}(k) + \exp(\sin(x_{1}(k))\sin(x_{2}(k)))\,u(k), \\
    \Delta^{\alpha} x_{2}(k+1) & = 0.035311\,x_{1}(k) + 0.001815\,x_{2}(k) + \left(1 + (x_{1}(k)+x_{2}(k))^2\right) u(k)
\end{array}
\right.
\end{align}
We use system~\eqref{eq:ultra_capacitor} to generate the datasets required for dynamic inference over the domain $x \in [-1, 1] \times [-0.3, 0.3]$, with the fractional order parameter set to $\alpha = 0.2$. Accordingly, the drift vector and control vector fields are given by:
\begin{align*}
    f(x(k)) & = \begin{bmatrix}
        x_{2}(k) \\
        0.035311\,x_{1}(k) + 0.001815\,x_{2}(k)
    \end{bmatrix}, \\
    g(x(k)) & = \begin{bmatrix}
        \exp(\sin(x_{1}(k))\sin(x_{2}(k))) \\
        1 + (x_{1}(k)+x_{2}(k))^2
    \end{bmatrix}.
\end{align*}
In these simulations, we use $L = 5$ expansion terms in the truncated orthonormal basis series. The numerical results are illustrated in Figure~\ref{fig:capacitor_DVF}. To assess robustness, we also include results obtained under measurement noise, following the noise model introduced in Remark~\ref{rem:noise}.

\subsection{Comparison of Fractional-Order and Integer-Order Systems}
In this section, we demonstrate the advantages of using fractional-order systems for modeling non-Markovian processes. To generate non-Markovian signals, we use the systems defined in~\eqref{eq:logistic_map}-\eqref{eq:lotka-volterra} to create the necessary datasets. Our proposed algorithms are then applied to identify corresponding fractional-order systems. Additionally, we employ a data-driven system identification algorithm for integer-order systems introduced in~\cite{narayanan2020disentangling}, using the same datasets.

\begin{figure}[!h]
\centering
\includegraphics[width=0.48\columnwidth]{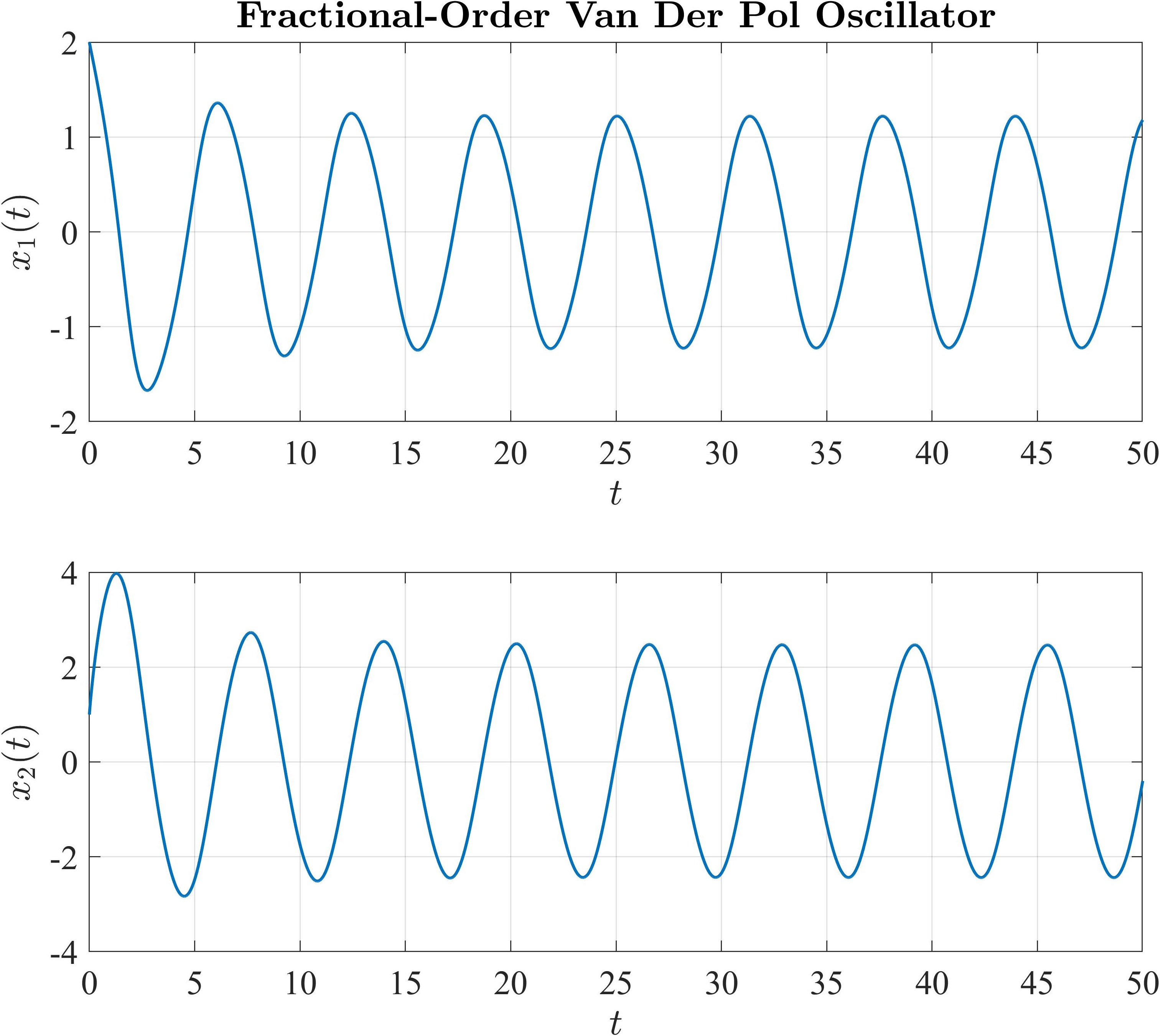} \hfill
\includegraphics[width=0.48\columnwidth]{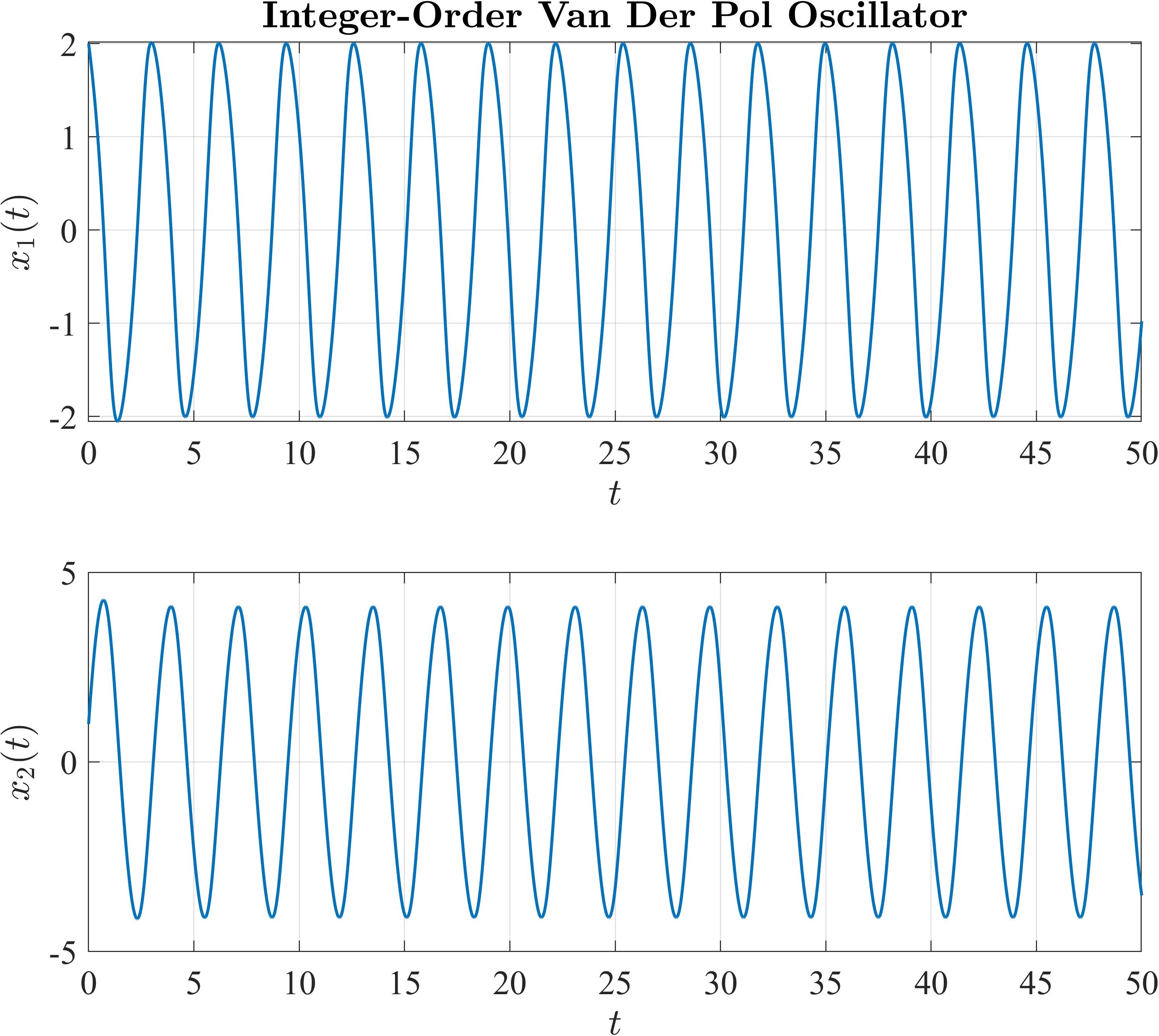}
\caption{Comparison of the System Response for Identified Fractional-Order Van der Pol Oscillator with the Order of $\alpha = 0.9$ and Integer-Order Van der Pol Oscillator.}
\label{fig:vanderpol_fractional_system_response_1}
\end{figure}

\begin{figure}[!h]
\centering
\includegraphics[width=0.48\columnwidth]{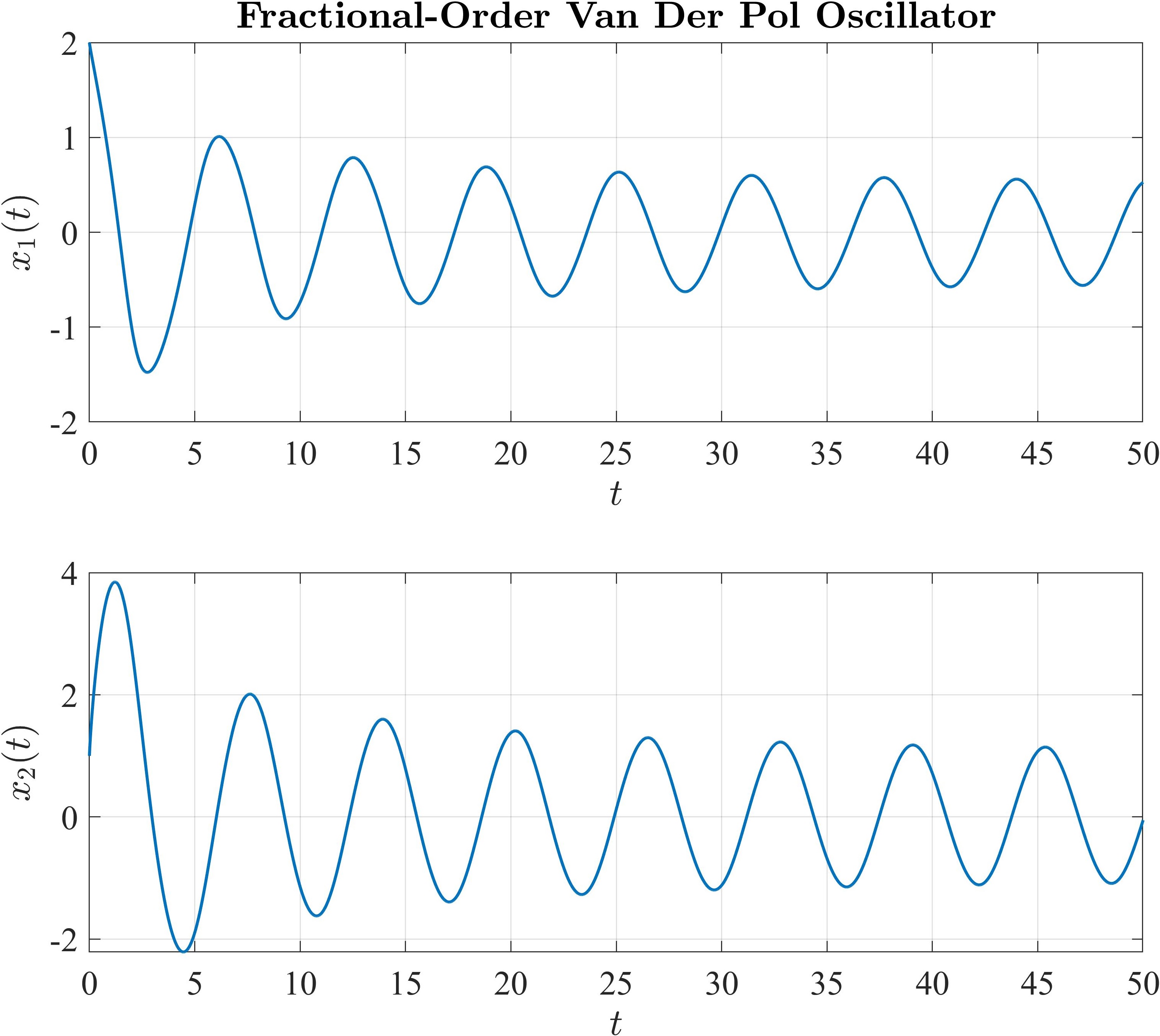}
\includegraphics[width=0.48\columnwidth]{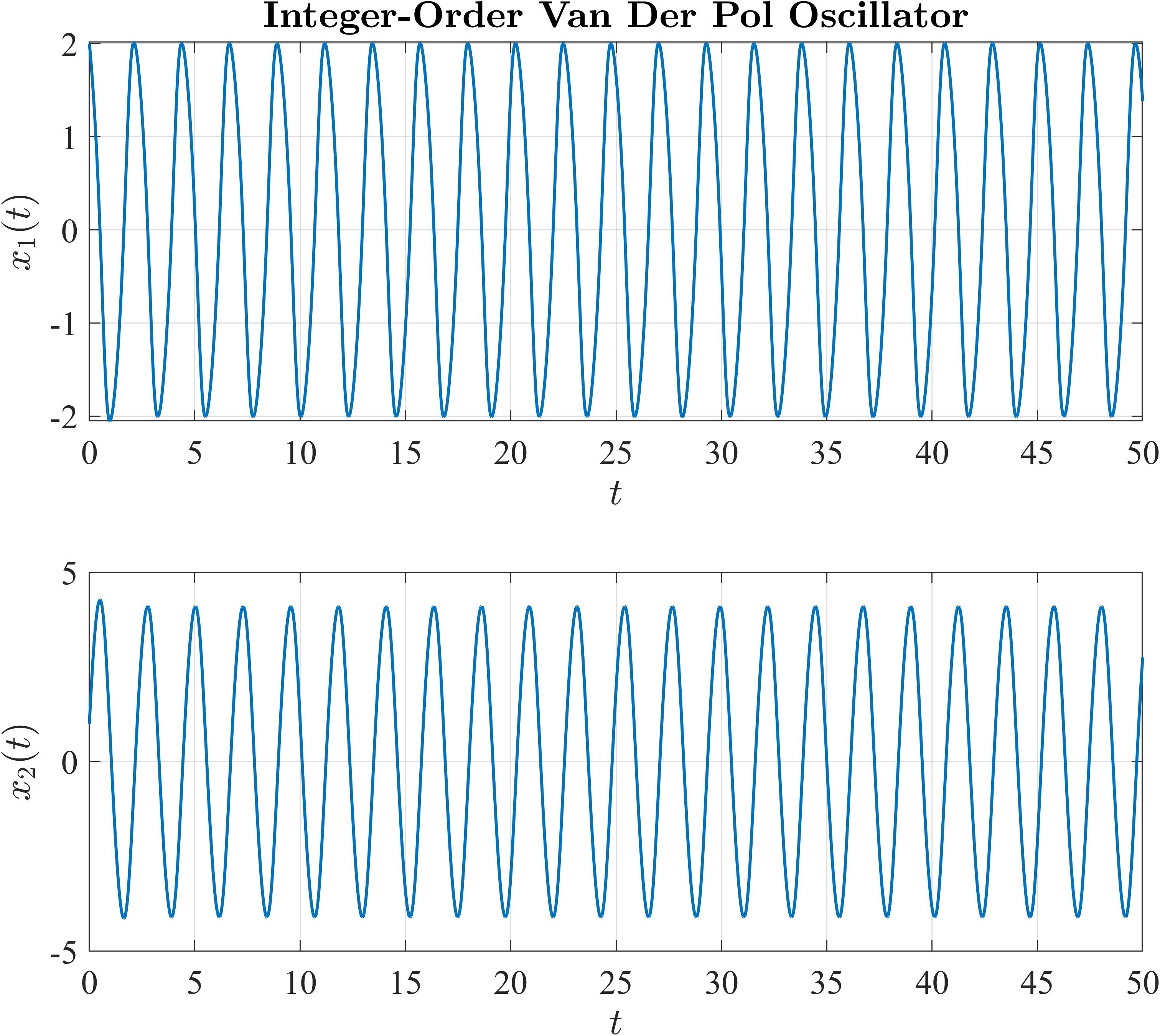}
\caption{Comparison of the System Response for Identified Fractional-Order Van der Pol Oscillator with the Order of $\alpha = 0.85$ and Integer-Order Van der Pol Oscillator.}
\label{fig:vanderpol_fractional_system_response_2}
\end{figure}

\begin{figure}[!h]
\centering
\includegraphics[width=0.48\columnwidth]{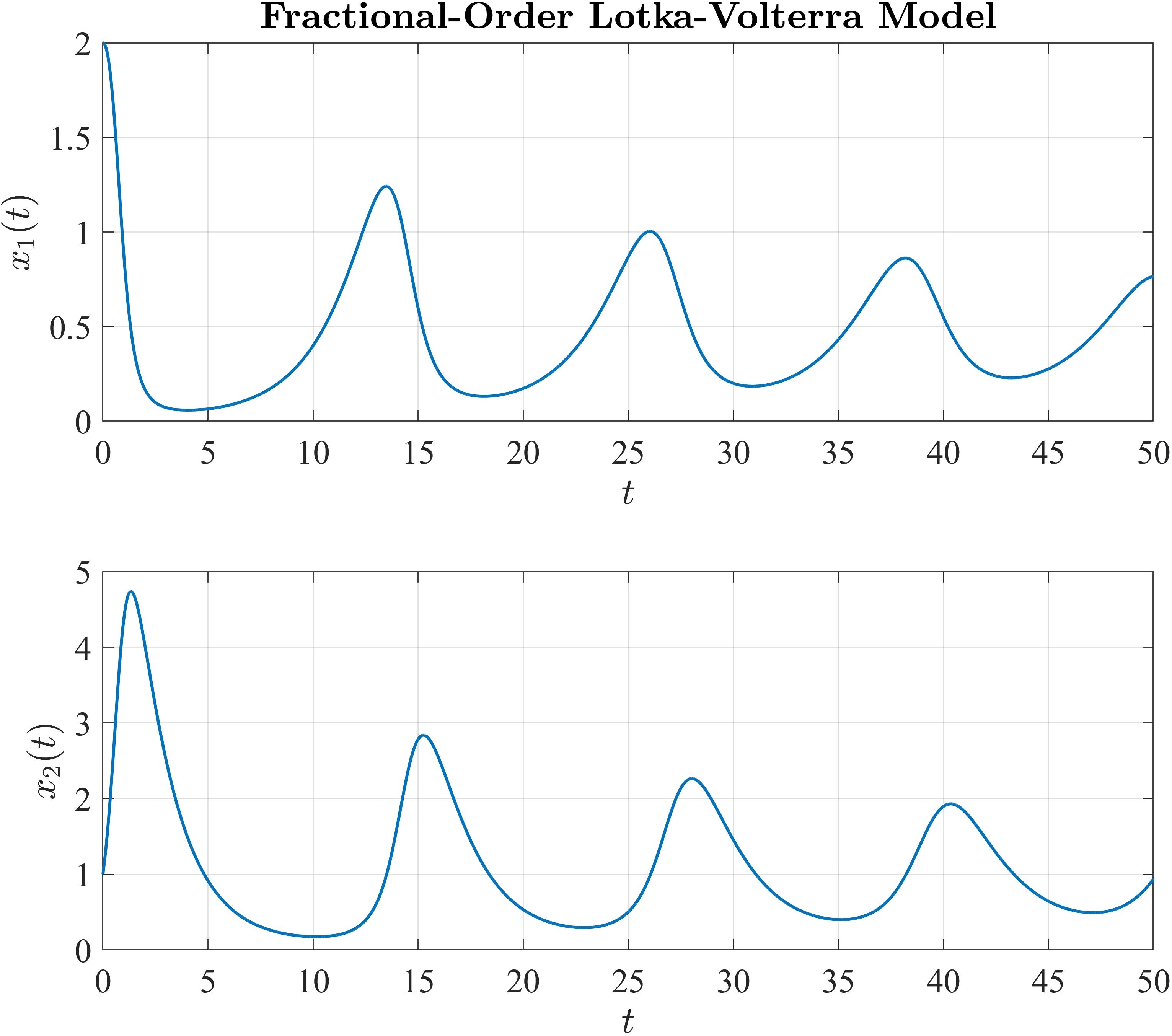} \hfill
\includegraphics[width=0.48\columnwidth]{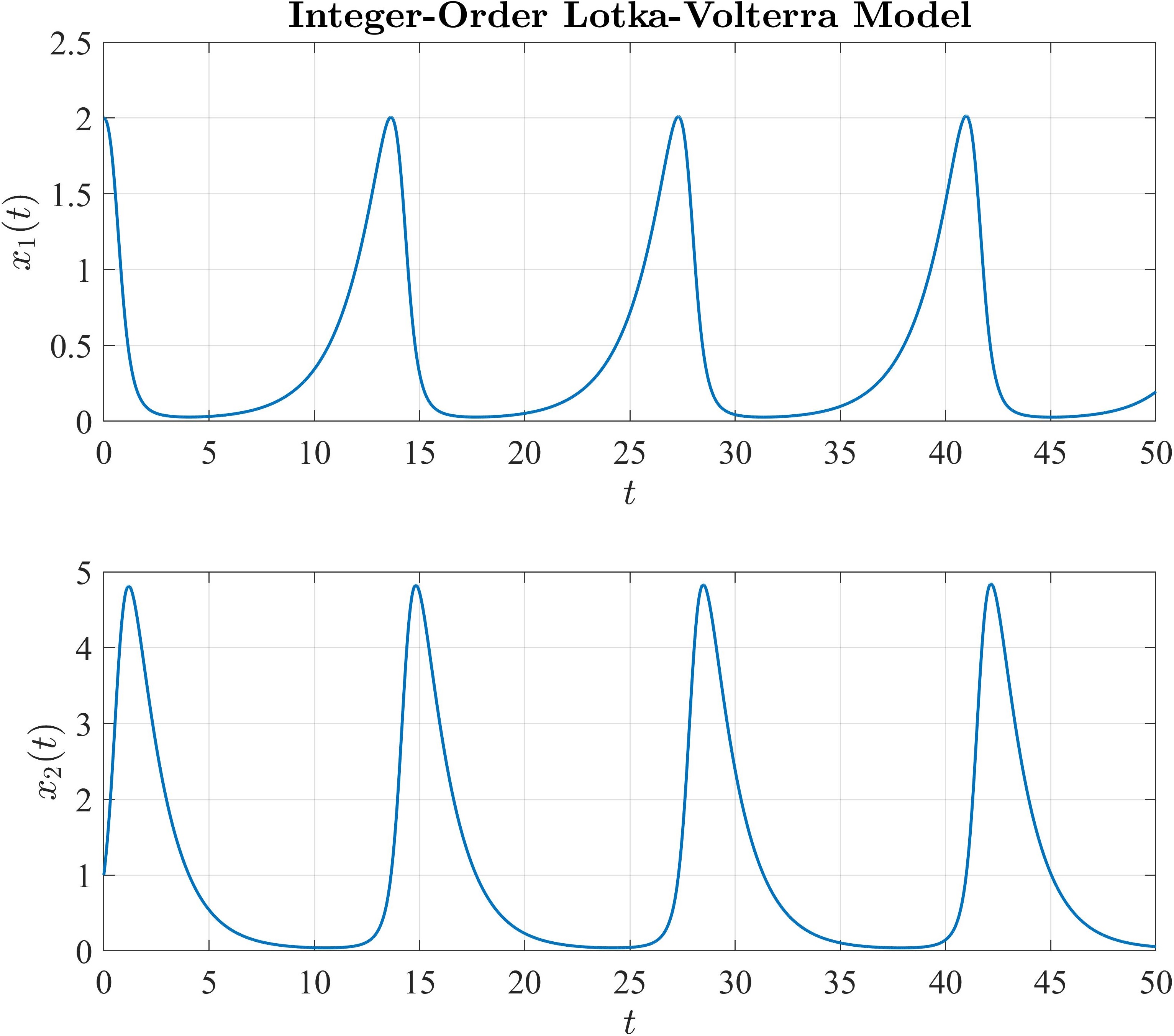}
\caption{Comparison of the System Response for Identified Fractional-Order Lotka-Volterra Model with the Order of $\alpha = 0.98$ and Integer-Order Lotka-Volterra Model.}
\label{fig:lotka_fractional_system_response_1}
\end{figure}

\begin{figure}[!h]
\centering
\includegraphics[width=0.48\columnwidth]{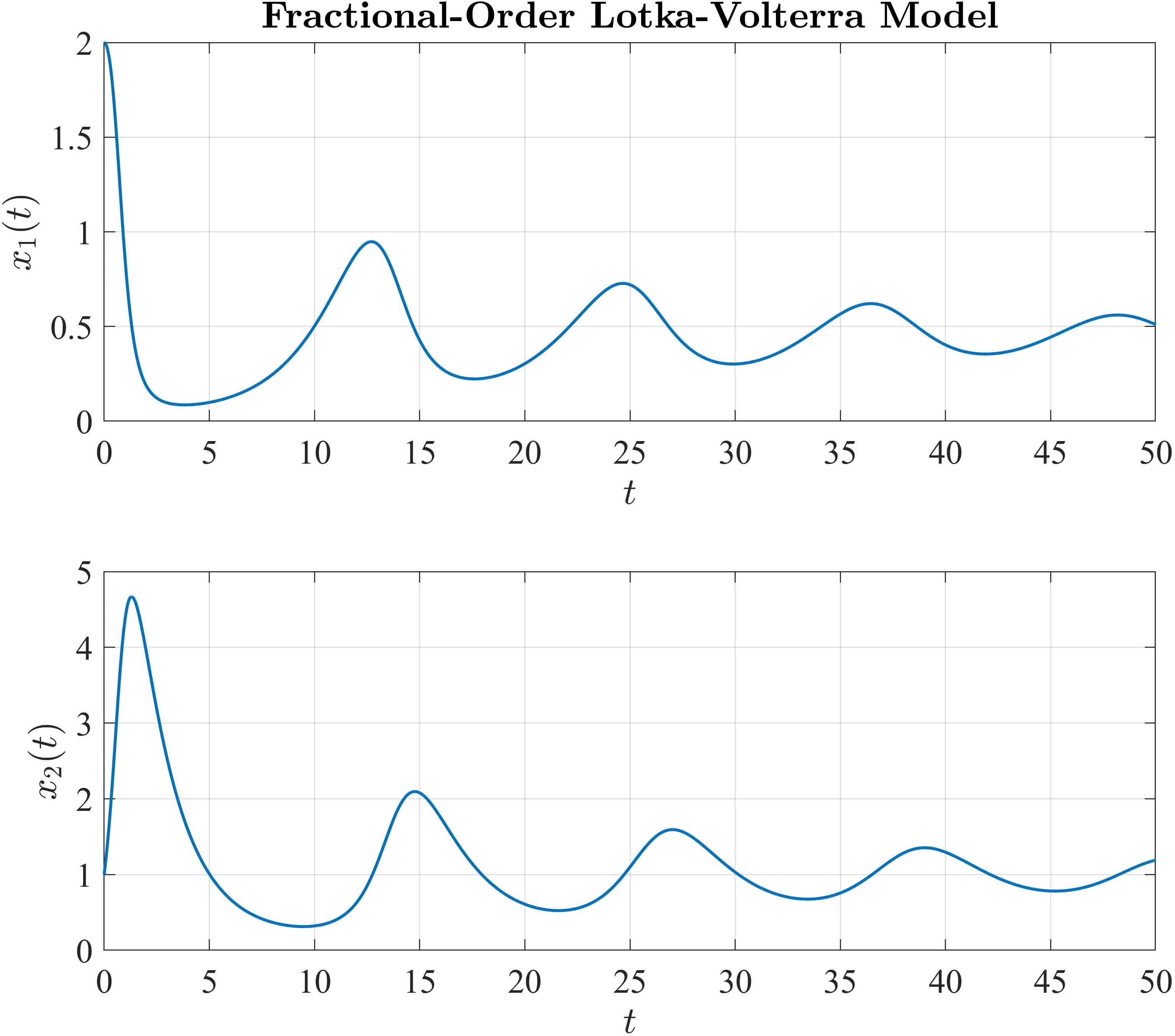} \hfill
\includegraphics[width=0.48\columnwidth]{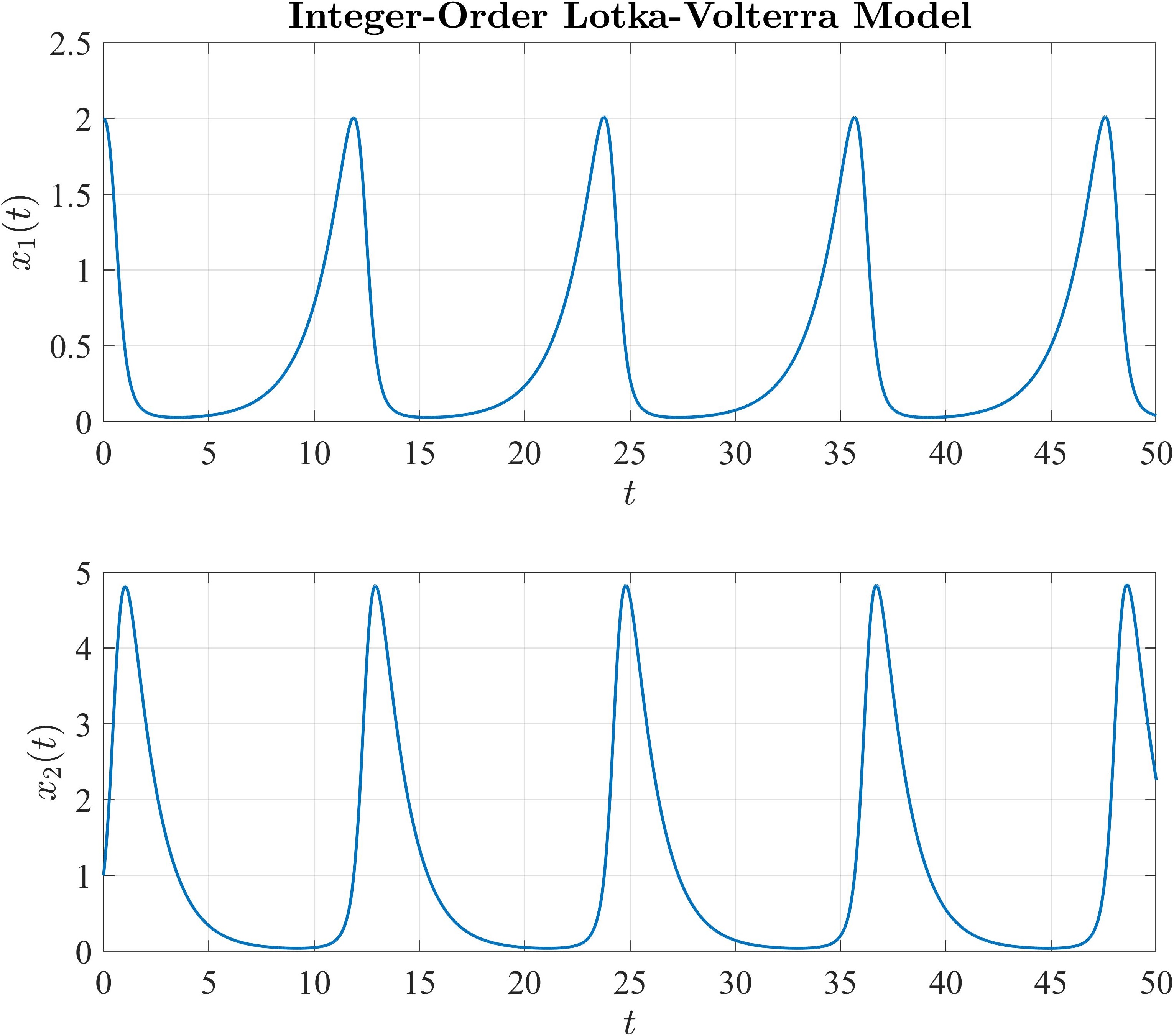}
\caption{Comparison of the System Response for Identified Fractional-Order Lotka-Volterra Model with the Order of $\alpha = 0.96$ and Integer-Order Lotka-Volterra Model.}
\label{fig:lotka_fractional_system_response_2}
\end{figure}

\begin{figure}[!h]
\centering
\includegraphics[width=0.96\columnwidth]{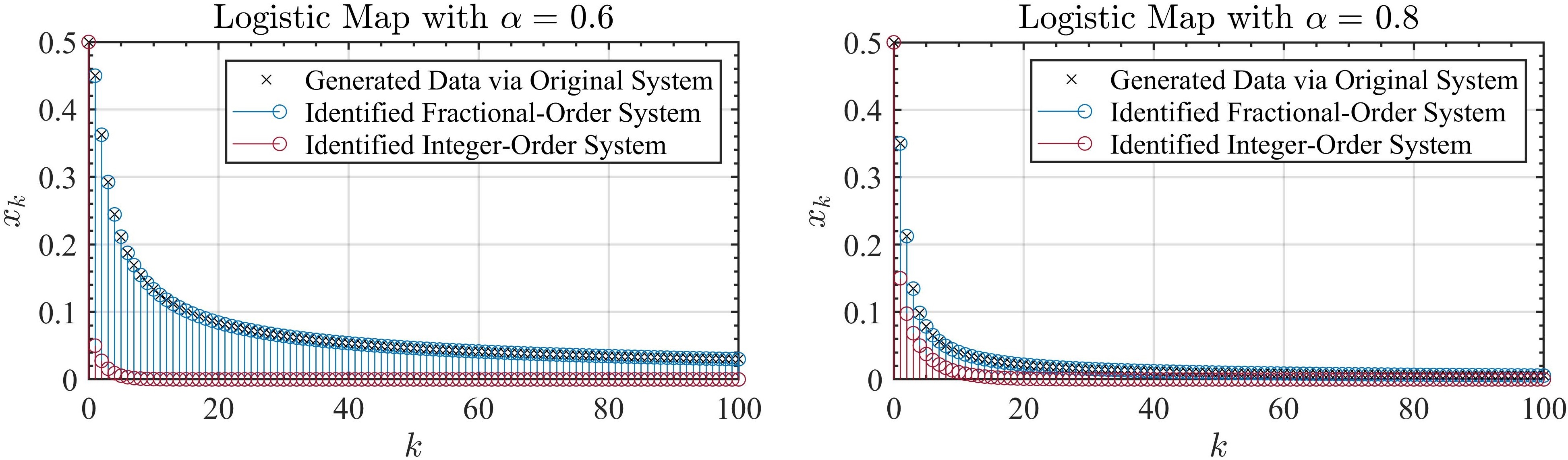}
\caption{Comparison of the System Response for Identified Fractional-Order Logistic and Integer-Order Logistic Map.}
\label{fig:logistic_system_response}
\end{figure}

\begin{figure}[!h]
\centering
\includegraphics[width=0.48\columnwidth]{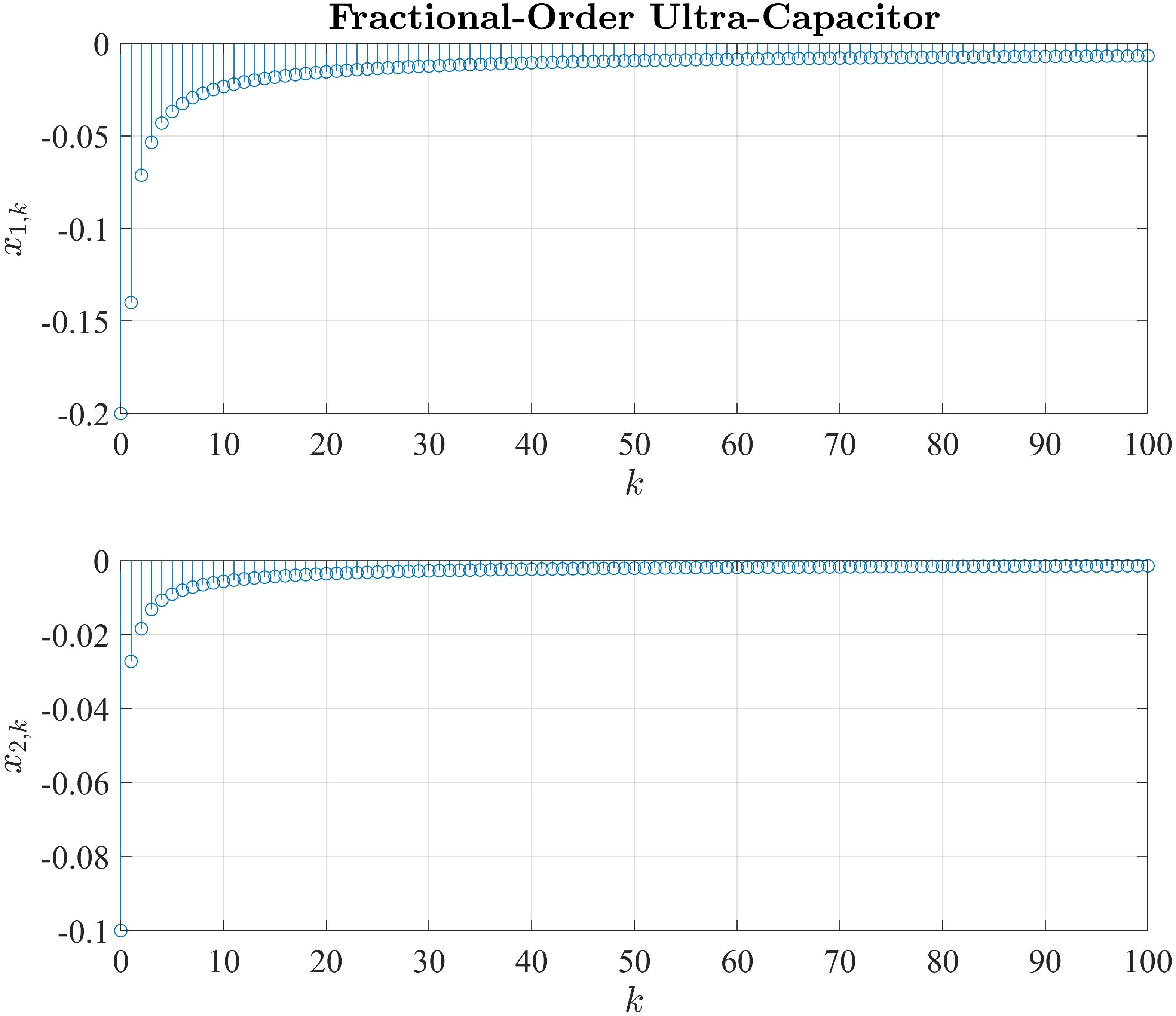}
\includegraphics[width=0.48\columnwidth]{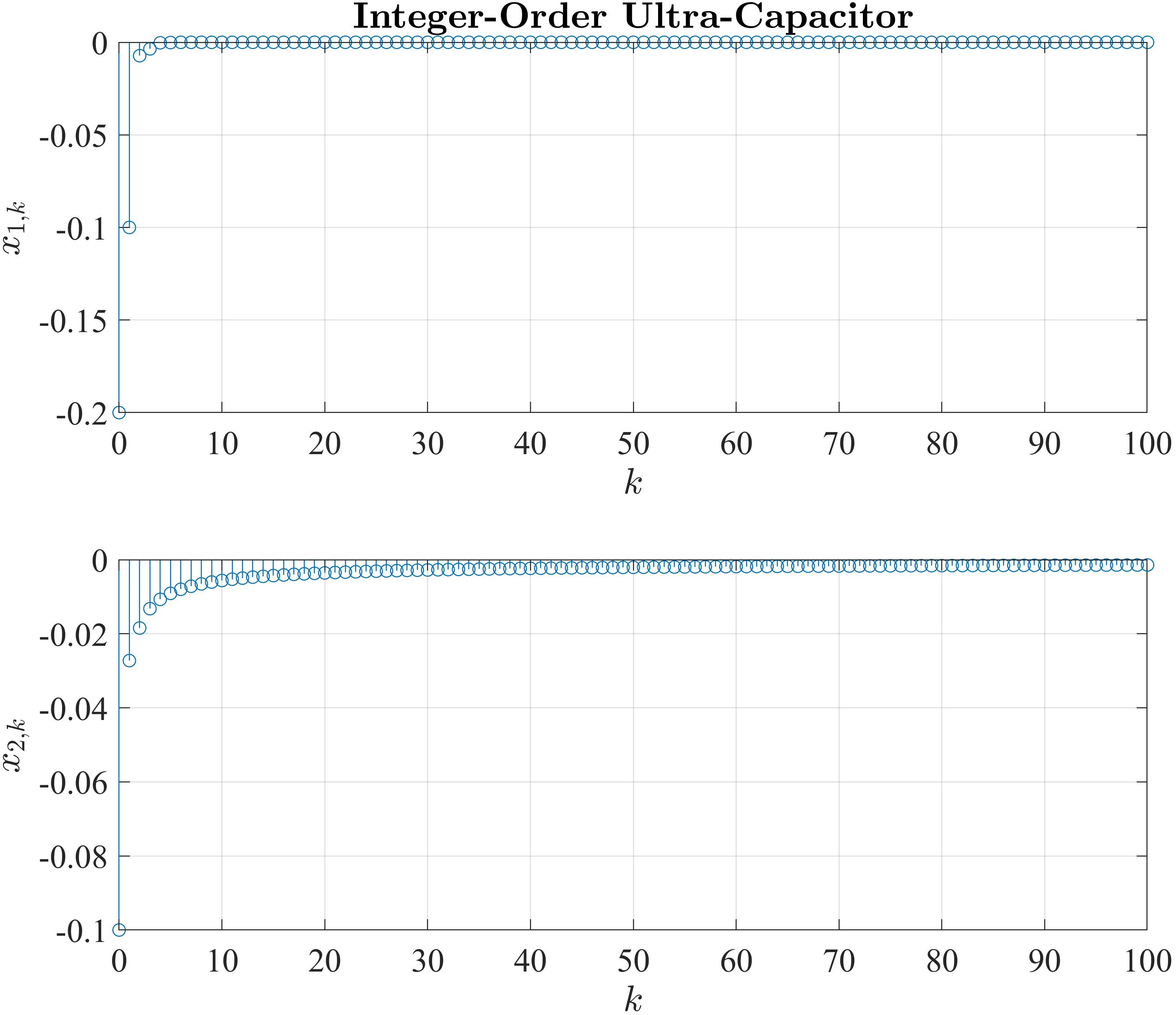}
\caption{Comparison of the System Response for Identified Fractional-Order Lotka-Volterra Model with the Order of $\alpha = 0.3$ and Integer-Order Lotka-Volterra Model.}
\label{fig:capacitor_fractional_system_response_1}
\end{figure}

To evaluate the impact of system order on both the system response and memory dependent behavior, we compare the responses of the inferred fractional-order and integer-order systems. The system trajectories obtained from the identified models are illustrated in Figures~\ref{fig:vanderpol_fractional_system_response_1}-\ref{fig:capacitor_fractional_system_response_1}. These simulations clearly show that integer-order models fail to accurately capture the dynamics of non-Markovian systems. Moreover, the approximation error increases as the fractional order~$\alpha$ decreases, reflecting stronger memory dependence—consistent with observations in~\cite{gupta2021non, flandrin1992wavelet}.

\section{Conclusion}\label{section:conclusion}
This paper presents data-integrated frameworks for learning the dynamics of fractional-order nonlinear systems. The proposed algorithms consist of two main steps. In the first step, input-output experiments are designed to generate datasets that capture the system's behavior. In the second step, the collected data are used to learn the system dynamics by approximating the drift-vector and control-vector fields. These fields are modeled using truncated series of orthonormal basis functions.
Theoretical developments are supported by numerical simulations, which show that fractional-order systems more effectively capture long-range dependence and non-Markovian behavior due to their inherent memory-dependent properties.

As future work, we aim to investigate the impact of measurement noise on the performance of the proposed learning framework and to develop more robust algorithms. In particular, we plan to incorporate the Hurst exponent as a metric for assessing memory in time-series data.

\appendix
\section{Experimental Results (cont.)}\label{section:experiments_cont}
In this appendix, we present the approximation errors associated with the drift-vector and control-vector fields for the systems analyzed in Section~\ref{section:simulation}. These results quantitatively assess the accuracy of the learned dynamics by comparing the estimated functions against the true underlying models.

\begin{figure}[!h]
\centering
\includegraphics[width=0.48\columnwidth]{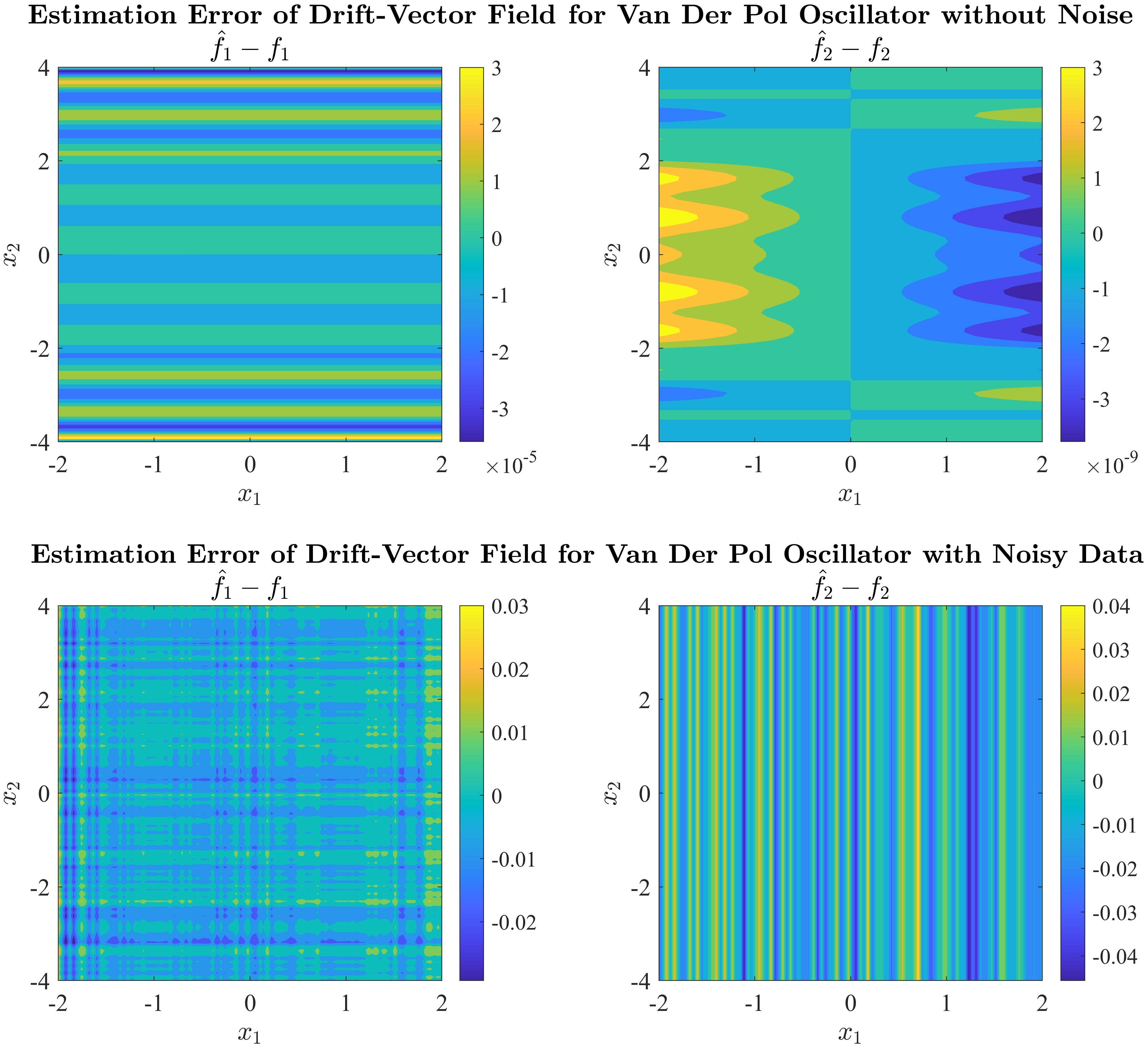} \hfill
\includegraphics[width=0.48\columnwidth]{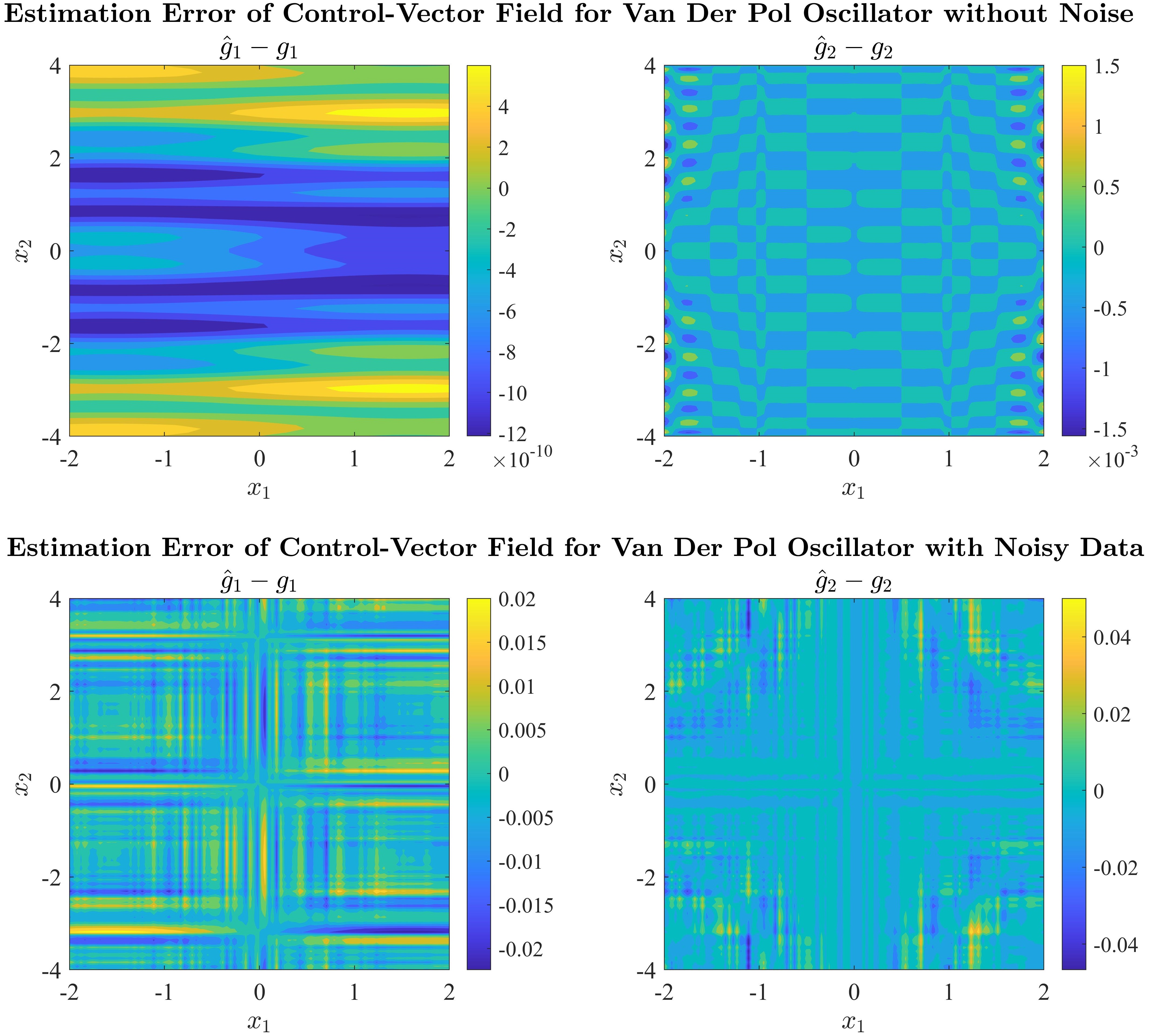}
\caption{Estimation Error of the Drift-Vector and Control-Vector Fields for Van der Pol oscillator.}
\label{fig:vanderpol_DVF_error}
\end{figure}

\begin{figure}[!h]
\centering
\includegraphics[width=0.48\columnwidth]{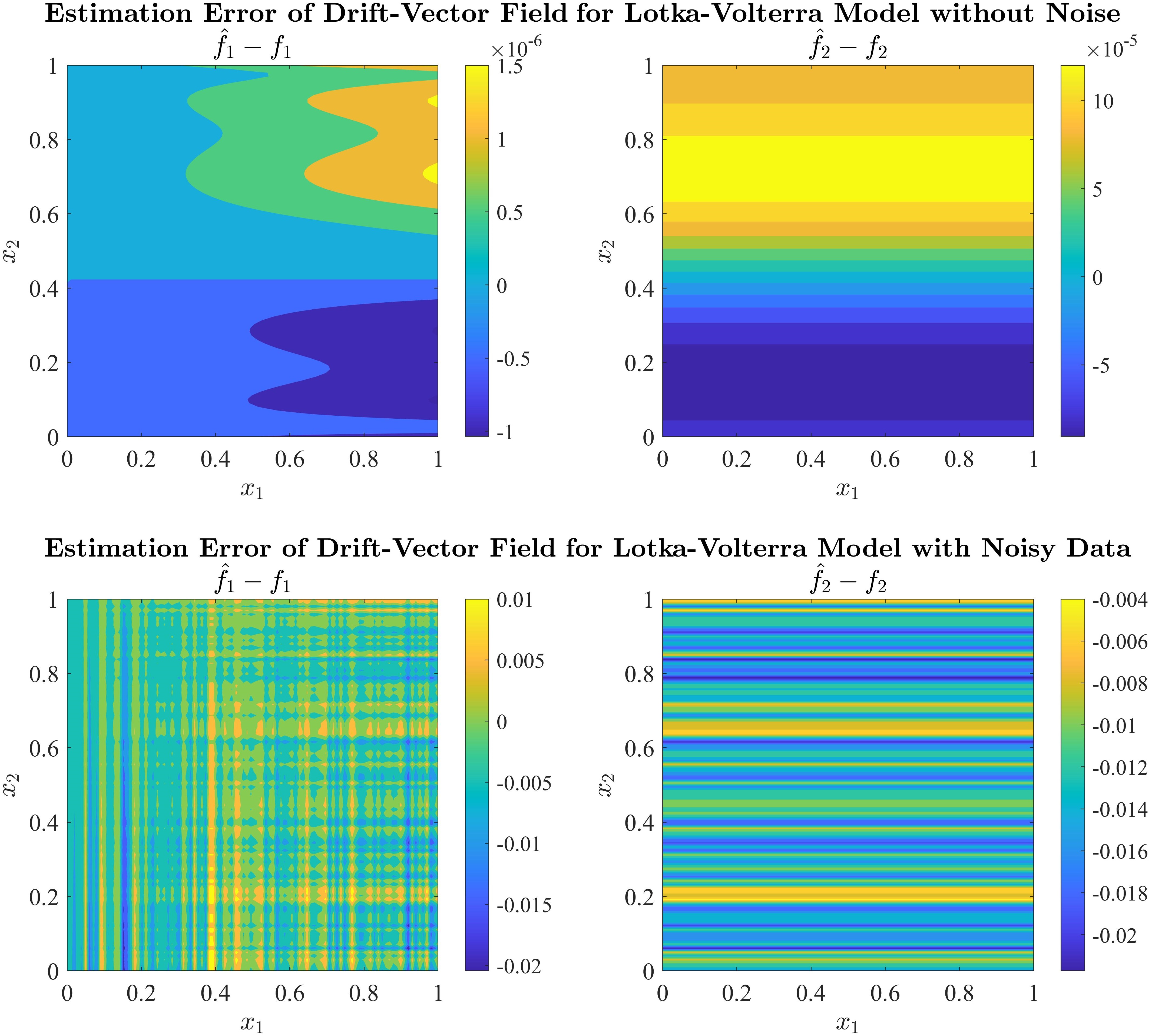} \hfill
\includegraphics[width=0.48\columnwidth]{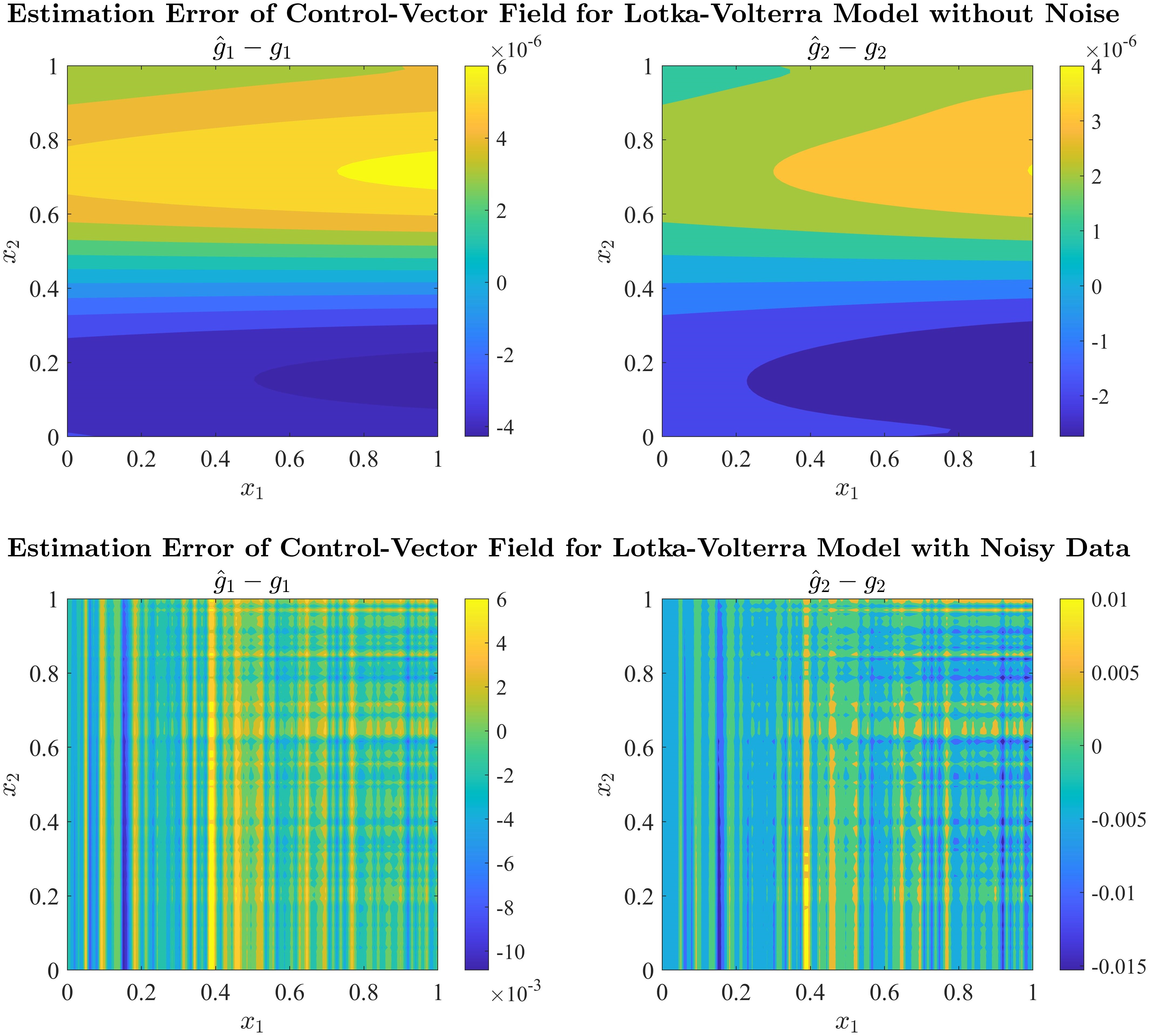}
\caption{Estimation Error of the Drift-Vector and Control-Vector Fields for Lotka-Volterra Model.}
\label{fig:lotka_DVF_error}
\end{figure}

\begin{figure}[!h]
\centering
\includegraphics[width=0.48\columnwidth]{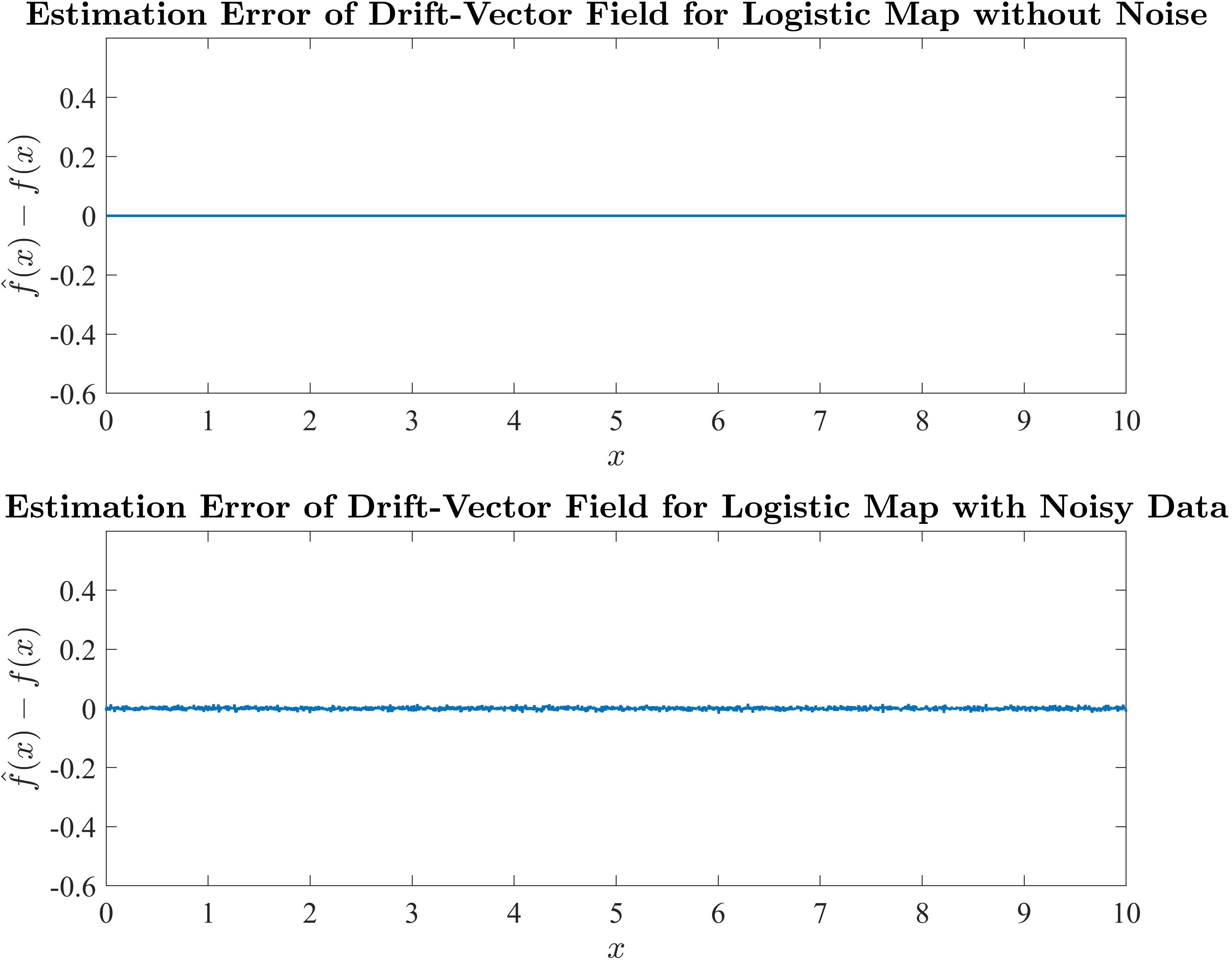} \hfill
\includegraphics[width=0.48\columnwidth]{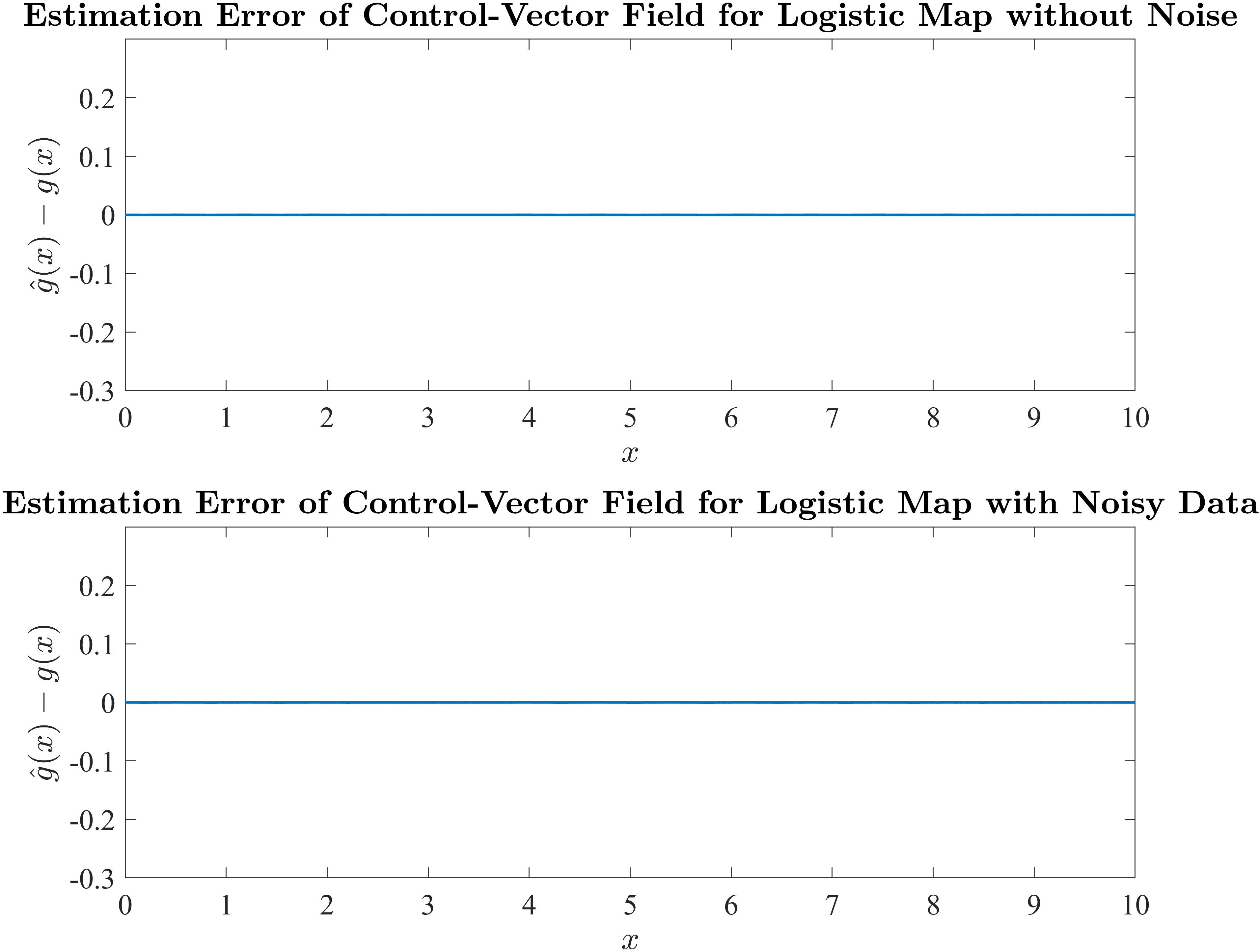}
\caption{Estimation Error of the Drift-Vector and Control-Vector Fields for Logistic Map.}
\label{fig:logistic_DVF_error}
\end{figure}

\begin{figure}[!h]
\centering
\includegraphics[width=0.48\columnwidth]{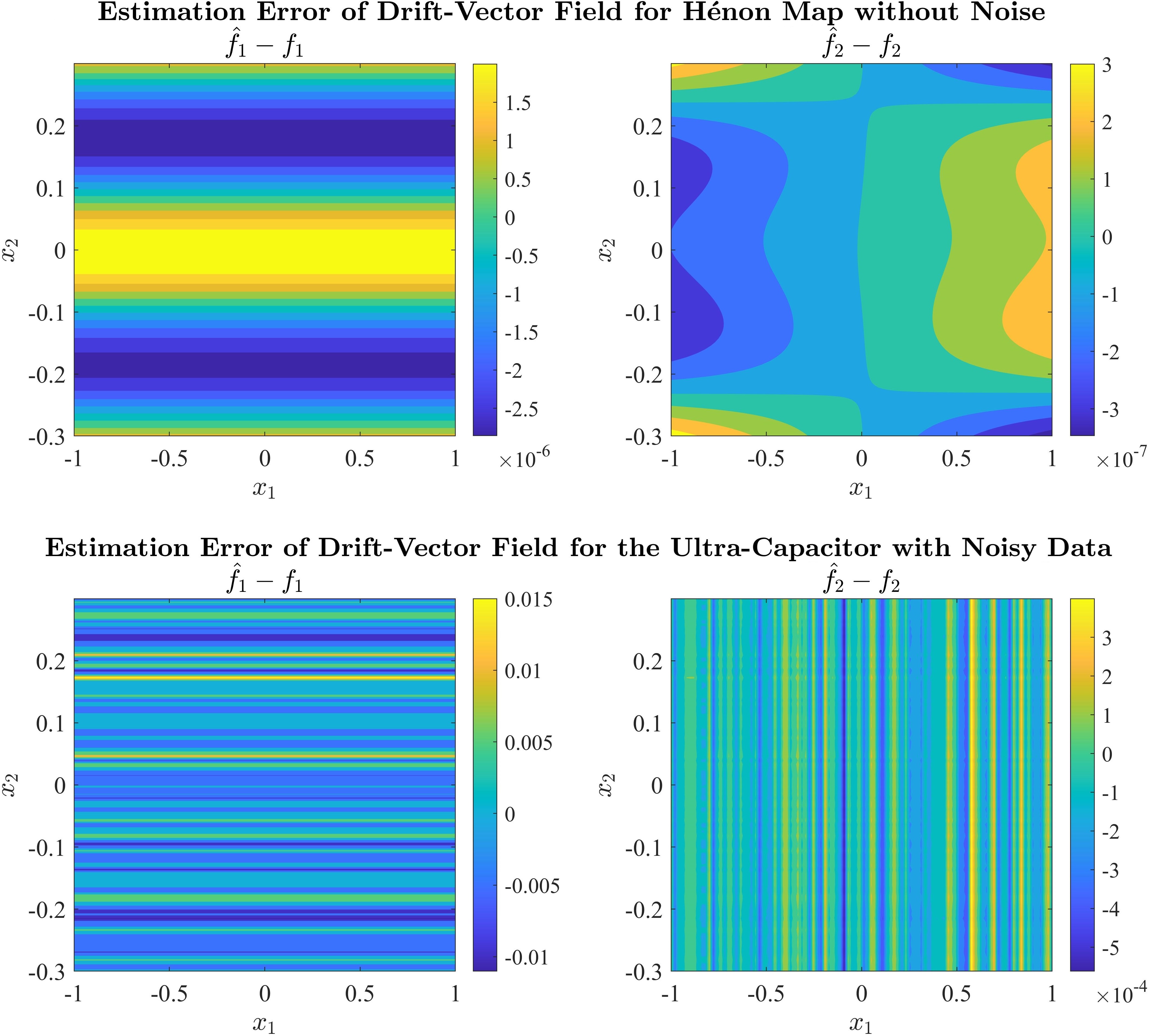} \hfill
\includegraphics[width=0.48\columnwidth]{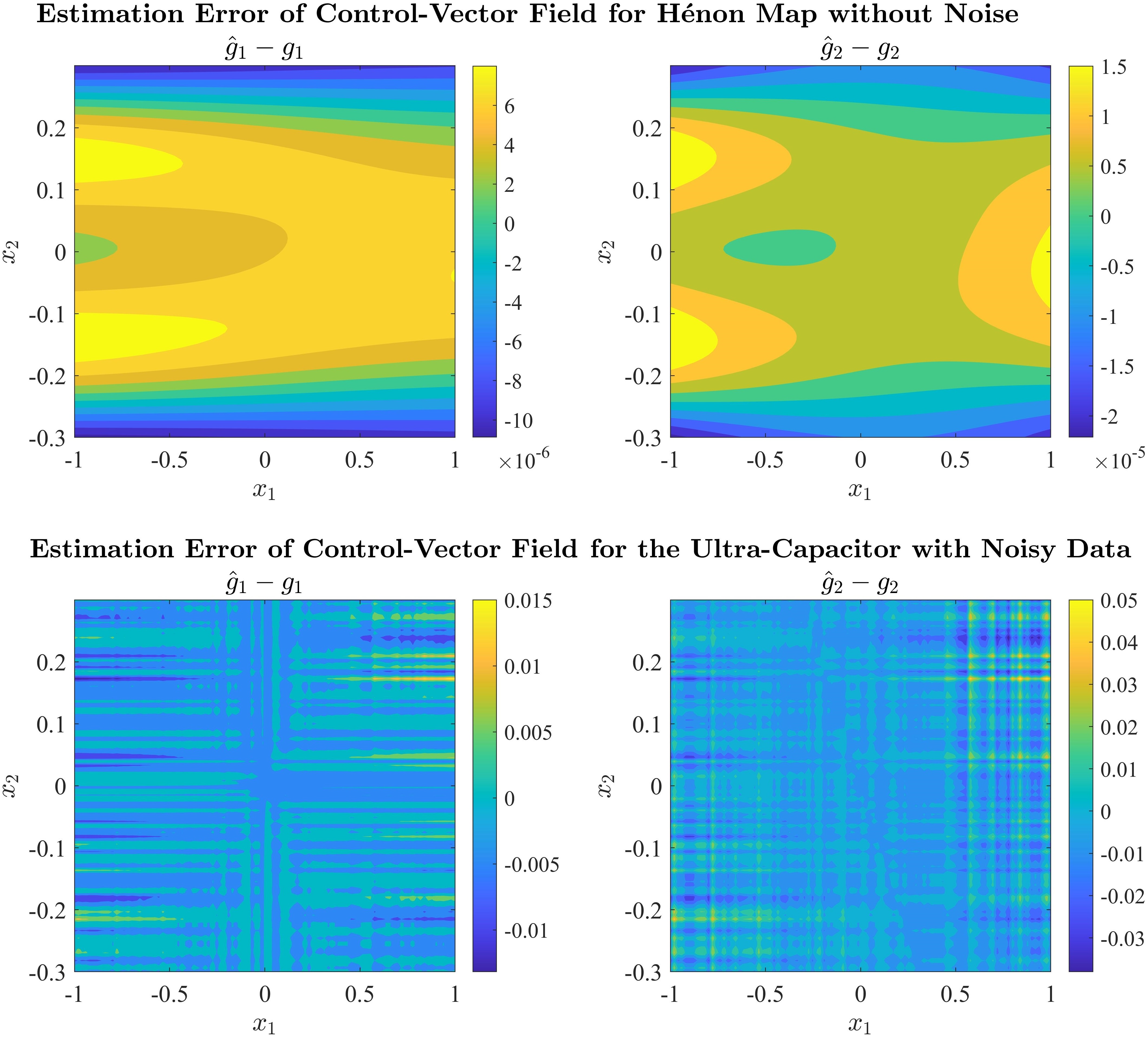}
\caption{Estimation Error of the Drift-Vector and Control-Vector Fields for Ultra-Capacitor.}
\label{fig:capacitor_DVF_error}
\end{figure}


\bibliographystyle{siamplain}
\bibliography{references}

\end{document}

%% file: ex_shared.tex
\usepackage{lipsum}
\usepackage{amsfonts}
\usepackage{graphicx}
\usepackage{epstopdf}
\usepackage{algorithmic}
\ifpdf
  \DeclareGraphicsExtensions{.eps,.pdf,.png,.jpg}
\else
  \DeclareGraphicsExtensions{.eps}
\fi

\usepackage{enumitem}
\setlist[enumerate]{leftmargin=.5in}
\setlist[itemize]{leftmargin=.5in}

\usepackage{lmodern}
\usepackage{kpfonts}
\usepackage{xcolor}
\usepackage{soul} 

\newcounter{subsubsubsection}[subsubsection]
\renewcommand\thesubsubsubsection{\thesubsubsection.\arabic{subsubsubsection}}
\newcommand\subsubsubsection[1]{%
  \refstepcounter{subsubsubsection}%
  \paragraph{\thesubsubsubsection\ #1}%
}

\newtheorem{problem}{Problem}
\newtheorem{notation}{Notation}

\newcommand{\bR}{\mathbb{R}}

\newcommand{\cX}{\mathcal{X}}

\newsiamremark{remark}{Remark}
\newsiamremark{hypothesis}{Hypothesis}
\crefname{hypothesis}{Hypothesis}{Hypotheses}
\newsiamthm{claim}{Claim}

\headers{A Data-Integrated Framework for Learning FO Nonlinear Dynamical Systems}{B. Yaghooti, C. Li, and B. Sinopoli}

\title{A Data-Integrated Framework for Learning Fractional-Order Nonlinear Dynamical Systems
}

\author{Bahram Yaghooti\thanks{Department of Electrical and Systems Engineering, Washington University in St. Louis, St. Louis, MO, USA 
  (\email{byaghooti@wustl.edu}, \email{lichengyu@wustl.edu}, \email{bsinopoli@wustl.edu}).}
  \and Chengyu Li\footnotemark[1]
\and Bruno Sinopoli\footnotemark[1]}

\usepackage{amsopn}

\makeatletter
\newcommand*{\addFileDependency}[1]{
  \typeout{(#1)}
  \@addtofilelist{#1}
  \IfFileExists{#1}{}{\typeout{No file #1.}}
}
\makeatother

\newcommand*{\myexternaldocument}[1]{%
    \externaldocument{#1}%
    \addFileDependency{#1.tex}%
    \addFileDependency{#1.aux}%
}
